\documentclass[%
 reprint,
 superscriptaddress,
 amsmath,amssymb,
 aps,
]{revtex4-2}

\usepackage{graphicx}
\usepackage{dcolumn}
\usepackage{bm}
\usepackage{float}
\usepackage{algorithm}
\usepackage{algpseudocode}
\usepackage{placeins}

\usepackage{titlesec}
\titleformat{\section}
  {\normalfont\normalsize\bfseries\centering}
  {\thesection.}
  {0.5em}
  {\MakeUppercase}

\titleformat{\subsection}
  {\normalfont\normalsize\bfseries\centering}
  {\thesubsection.}
  {0.5em}
  {}
  
\DeclareMathOperator{\sigmoid}{sigmoid}
\begin{document}
\raggedbottom

\preprint{APS/123-QED}

\title{Confidence-Aware Discriminative Readout of Neutral-Atom Arrays}
\author{Jaeick Bae}
\affiliation{Korea Research Institute of Standards and Science, Daejeon 34113, South Korea}
\affiliation{
Graduate School of Quantum Science and Technology, Korea Advanced Institute of Science and Technology, Daejeon 34141, South Korea
}

\author{Kangheun Kim}
\affiliation{Korea Research Institute of Standards and Science, Daejeon 34113, South Korea}
\author{Yunheung Song}
\affiliation{Korea Research Institute of Standards and Science, Daejeon 34113, South Korea}
\author{Junseo Ha}
\affiliation{Korea Research Institute of Standards and Science, Daejeon 34113, South Korea}
\author{Jeong Ho Han}
\affiliation{Korea Research Institute of Standards and Science, Daejeon 34113, South Korea}

\author{Jaewook Ahn}
\affiliation{
Graduate School of Quantum Science and Technology, Korea Advanced Institute of Science and Technology, Daejeon 34141, South Korea
}
\affiliation{Department of Physics, Korea Advanced Institute of Science and Technology, Daejeon 34141, South Korea}

\author{Hyunsoo Ha}
\affiliation{Center for Theoretical Physics—a Leinweber Institute, Massachusetts Institute of Technology, Cambridge, Massachusetts 02139, USA}
\author{Soonwon Choi}
\affiliation{Center for Theoretical Physics—a Leinweber Institute, Massachusetts Institute of Technology, Cambridge, Massachusetts 02139, USA}

\author{Jongchul Mun}
\email{jcmun@kriss.re.kr}
\affiliation{Korea Research Institute of Standards and Science, Daejeon 34113, South Korea}
\affiliation{
Graduate School of Quantum Science and Technology, Korea Advanced Institute of Science and Technology, Daejeon 34141, South Korea
}

\date{\today}

\begin{abstract}
Fast and reliable detection is a central requirement for neutral atom quantum computers, yet fluorescence imaging remains a limiting step for rapid experimental cycling, large-scale array operation, and quantum-error-correction protocols. In this work,  we demonstrate confidence-aware readout of \(^{171}\mathrm{Yb}\) atoms in optical tweezers using a discriminative classifier. Rather than forcing each atom-site image directly into a hard binary outcome, the proposed readout produces a measurement reliability score. This score is produced by a logistic-regression classifier trained on features derived from both the total photon count and the spatial pattern of fluorescence at each site. The resulting classifier achieves the same readout fidelity with approximately half the imaging time. At an exposure of 10 ms, it achieves \(99.83\%\) readout fidelity, comparable to the \(99.82\%\) achieved by conventional photon-count thresholding at 20 ms.  Importantly, the classifier score enables low-confidence outcomes to be converted into measurement-level erasures, leaving fewer residual unflagged errors at a fixed erasure budget. This
improvement is further accompanied by strong false-positive suppression, which
is important for downstream operations such as atom rearrangement where false
positives can create target defects. Monte Carlo rearrangement simulations show that
the reduced false-positive rate and confidence-based erasure conversion improve
predicted defect-free assembly performance. This work establishes confidence-aware readout using a discriminative classifier as a practical paradigm for fast and scalable neutral-atom quantum computing.
\end{abstract}

\maketitle

\section{Introduction}

Among current quantum computing platforms, neutral atom architectures have emerged as promising because of their scalability and flexible connectivity~\cite{bluvstein24_logical, manetsch25_tweezer6100}. Nevertheless, they still face several important systems-level limitations, including relatively slow atom transport~\cite{cicali25_transport, hwang25_transport} and fluorescence-based measurement times that remain long compared with those of physical qubit operations~\cite{saffman25}. This measurement overhead directly constrains the cycle time of neutral atom processors. It is especially important for repeated mid-circuit measurements in fault-tolerant circuits~\cite{bluvstein26, deist22}, rapid subarray rearrangement and state preparation in continuous-loading architectures~\cite{li25, chiu25}, and the suppression of idle errors during quantum error correction (QEC) cycles~\cite{veroni24}. Large-scale algorithms may require continuous fault-tolerant operation over several days, while neutral atom-specific QEC and simulation architectures increasingly rely on sub-millisecond syndrome extraction, rearrangement and few round syndrome measurements~\cite{cain26,zhao26_ultrahighrate,ismail26_transversal,gidney21_factor2048rsa}. These applications demand reliable fluorescence readout on a timescale compatible with processor's gate operation.

The central challenge in fast fluorescence imaging is not only to maintain high binary readout fidelity with fewer detected photons, but also to preserve information about which readout decisions are reliable. Strong photon scattering can heat atoms and lead to loss, whereas reducing the measurement time decreases the photon budget and broadens the overlap between the atom-present and atom-absent signal distributions. However, conventional fluorescence readout compresses each region of interest (ROI) into a binary decision by thresholding scalar photon count values. When the readout photon budget is limited, some ROI measurements fall close to the decision boundary and therefore carry substantial uncertainty in their assigned label. Hard thresholding discards this uncertainty by returning only a binary outcome, losing information about the reliability of that outcome.

This motivates a confidence-aware readout approach, in which the readout output is not limited to a hard occupied/empty label, but also includes a scalar measure of decision reliability. The value of such \textit{soft} measurement information is well established in QEC theory and has been experimentally explored most directly in superconducting qubit processors, where analog readout records can be incorporated into decoders instead of being discarded by \textit{hard} thresholding~\cite{ali2024softinfo, hanisch2026softinformation}. Related decoding ideas have also been studied for neutral atom QEC models, suggesting that fluorescence readout uncertainty can affect logical performance~\cite{majaniemi2026softinformation}. In neutral atom fluorescence experiments, however, image information is commonly compressed into binary occupation before being passed to downstream operations such as rearrangement, atom replacement, or reinitialization.

\begin{figure}[!ht]
\centering

\includegraphics[
  width=\columnwidth,
  trim={0.55cm 4cm 0.2cm 3cm},
  clip
]{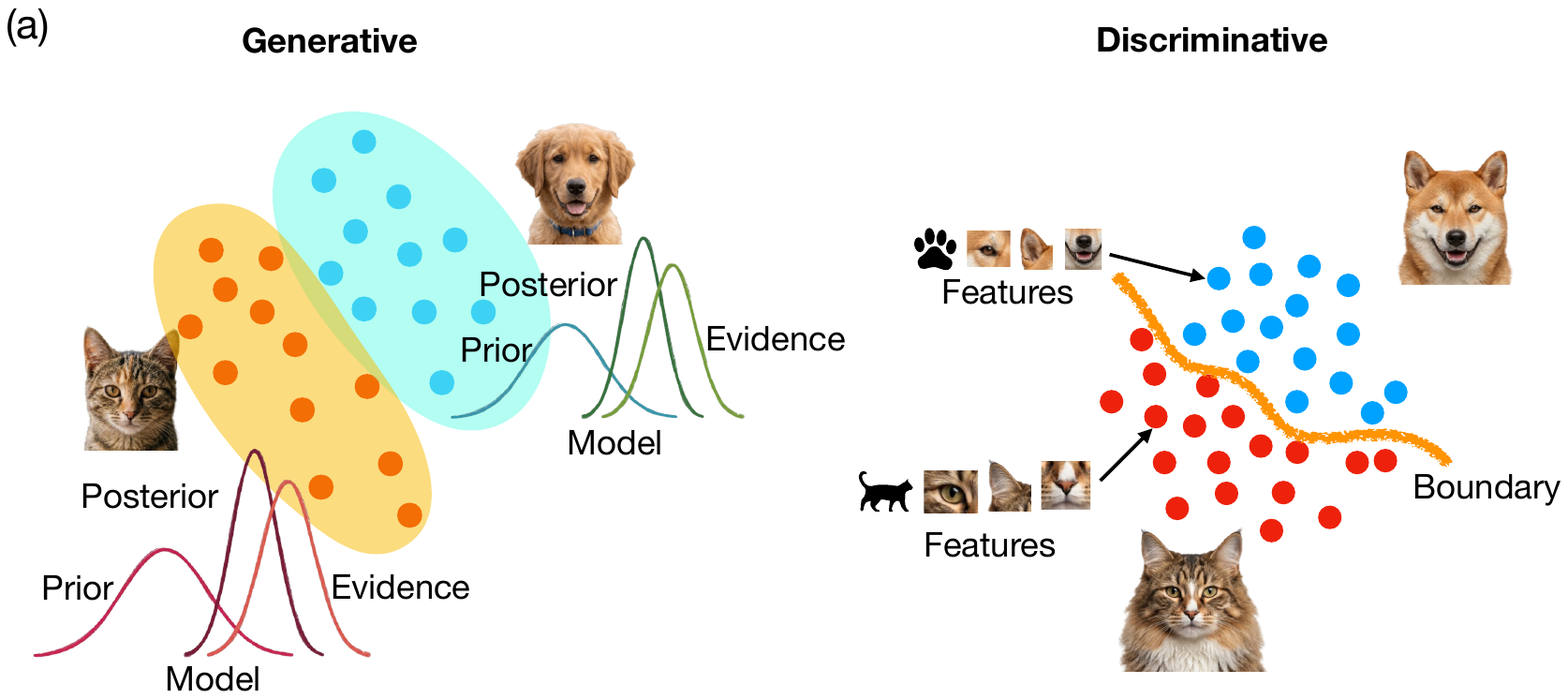}

\includegraphics[
  width=\columnwidth,
  trim={0.2cm 2.5cm 0cm 3cm},
  clip
]{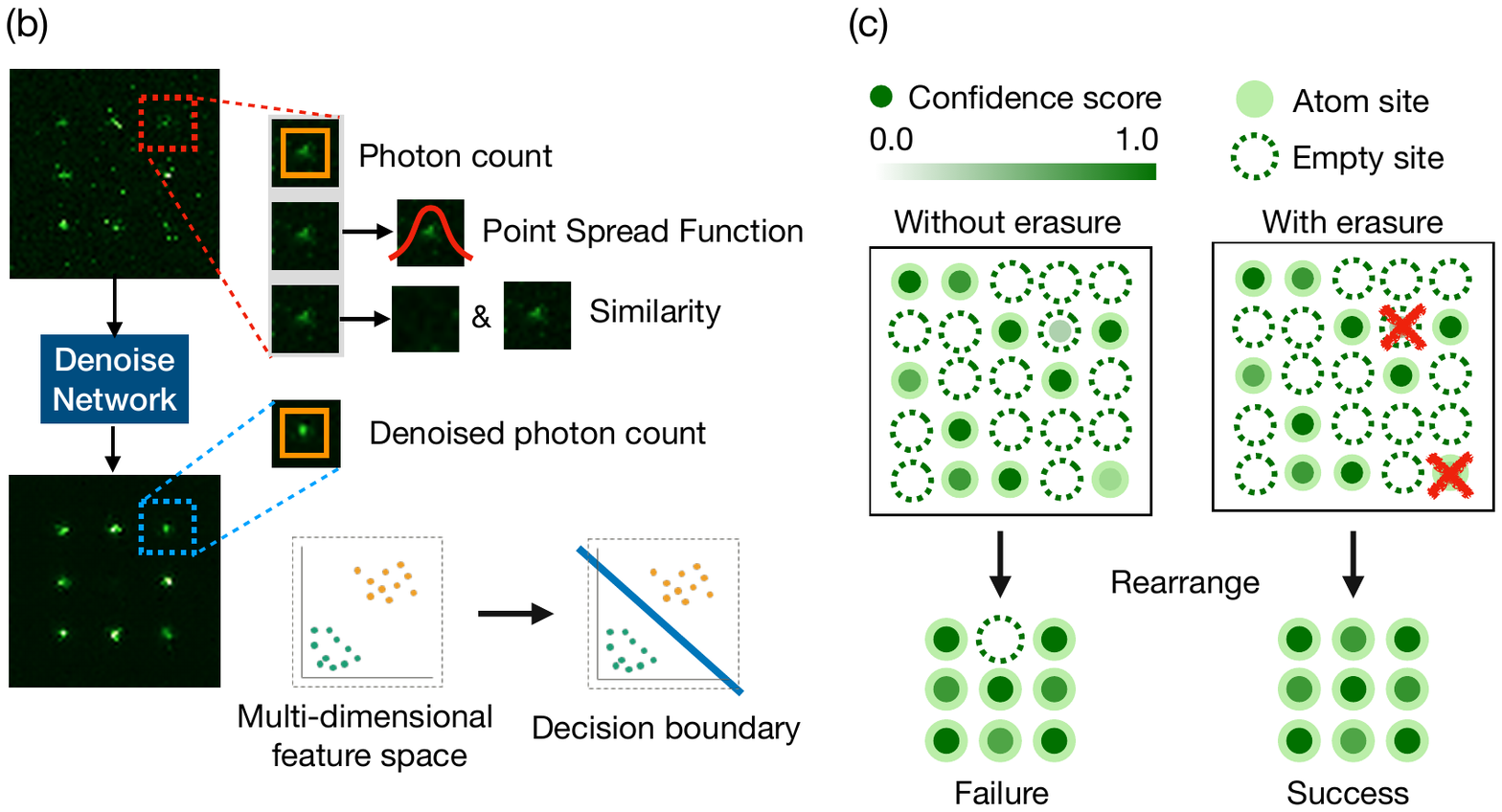}

\caption{\label{fig:overview}
\textbf{Overview of discriminative confidence-aware readout and its use in atom rearrangement.}
(a) Conceptual comparison between generative and discriminative classification.
A generative approach models how each class produces the measured data and
uses prior information and evidence to infer the posterior class probability.
A discriminative approach instead extracts features from the data and directly
learns a boundary that separates the two classes.
(b) Discriminative readout pipeline used in this work. From each fluorescence
image, we extract scalar region-of-interest (ROI) features including raw photon count, denoised
photon count, point-spread-function (PSF) template-matching score, and background-fitted camera-noise log-likelihood.
These features define a multi-dimensional feature space, in which
logistic regression learns the occupied/empty decision boundary.
(c) Use of the confidence score for measurement-erasure conversion in atom rearrangement.
Without erasure conversion, a low-confidence false-positive event can be treated as an occupied source site and lead to rearrangement failure.
With erasure-aware readout, the same low-confidence event is flagged and excluded from the source pool, enabling successful rearrangement.
}
\end{figure}

A key question is what kind of readout score provides useful reliability information for fluorescence readout images. To address this question, we consider two broad software-based classification: generative and discriminative methods [Fig.\ref{fig:overview}(a)]. A generative approach specifies how each class produces the observed data, for
example through class-conditional image likelihoods \(p(R|B)\) and \(p(R|D)\),
where \(R\) is the measured fluorescence image and \(B\) and \(D\) denote the
atom-present and atom-absent classes. Bayes' rule is then used to infer the posterior class
probability. This approach is physically interpretable and can
be data-efficient when the image-formation and noise models are accurate.
Nevertheless, its performance can be limited by mismatch between the analytical model
and the experimental image-distribution. A discriminative approach instead uses
labeled data to learn the decision boundary between atom-present and atom-absent images
directly. This can make the classifier less sensitive to misspecification of image-distribution details that are irrelevant to the classification task~\cite{ulusoy05}, and
its performance typically improves as more representative labeled training data
become available~\cite{ng01,raina03}. Although Bayesian inference yields the optimal decision rule when the
class-conditional generative model and class priors are correct~\cite{duda01},
accurate generative modeling can be difficult in experimental settings.
This motivates a discriminative readout strategy that learns the
empirical occupied/empty boundary from labeled data.

In this work, we demonstrate that a discriminative classifier's output value improves both fast readout and confidence-based error flagging for neutral atom fluorescence readout images. The classifier combines physics-derived and AI-learned features, including raw photon counts, photon counts after image noise reduction, empirical spatial information, and camera-noise likelihood features [Fig.\ref{fig:overview}(b)]. Relative to conventional thresholding and a Bayesian posterior-rule baseline, the resulting score improves the empirical occupancy readout fidelity and yields approximately a $2 \times$ reduction in the measurement time required to reach comparable readout fidelity. Beyond this improvement, the discriminative score provides more useful reliability information: when low-confidence outcomes are converted into measurement-level erasures, the classifier leaves fewer residual errors than the conventional thresholding or Bayesian baseline.

In particular, we introduce an AI-driven denoised-ROI feature that reduces class overlap and lowers false-positive rate (FPR), defined here as the probability that an atom-absent ROI is incorrectly classified as atom-present. When the denoised-ROI feature is incorporated into the multi-feature discriminative classifier, the resulting AI-assisted readout strongly suppresses false positives, reducing the FPR by more than one order of magnitude. Unlike generative image approaches such as Generative Adversarial Network (GAN)-based denoising~\cite{mude25}, our method does not require clean target images for denoiser training and suppresses hallucinations.

These gains are directly relevant to assembly of defect-free atomic qubit arrays. During rearrangement, false-positive occupancy errors are particularly detrimental. If an empty reservoir site is falsely identified as occupied, it may be selected as an atom source, leading to an attempted move from an empty trap and leaving a defect in the target array. Confidence-based erasure mitigates this failure case by excluding
ambiguous sites from the source pool before rearrangement [Fig.\ref{fig:overview}(c)]. Although increasing the readout exposure time improves classification fidelity, our simulations show that fast readout combined with confidence-based erasure can yield higher defect-free success rates than longer readout without erasure filtering.

\section{Overview of experimental setup}

The imaging method used in this work follows the dual-tone narrow-line imaging scheme described in the previous work~\cite{song26}. We provide a brief summary of the scheme and focus on the experimental modifications that were necessary for simultaneous acquisition of readout and reference images. We imaged ${}^{171}\mathrm{Yb}$ atoms trapped in a shallow 759.4~nm clock-magic tweezer using dual-tone driving on the $\left| {}^1S_0,\; F=1/2,\; m_F=\pm 1/2 \right\rangle $ and $ \left| {}^3P_1,\; F'=3/2,\; m_F'=\pm 1/2 \right\rangle$ intercombination transition at 556~nm. A $3 \times 3$ tweezer array with a beam waist of $0.9~\mu\mathrm{m}$ and a trap depth of $200~\mu\mathrm{K}$ was used. 

\begin{figure}[!ht]
\centering
\includegraphics[
  width=\columnwidth,
  trim={2cm 6cm 3cm 3cm},
  clip
]{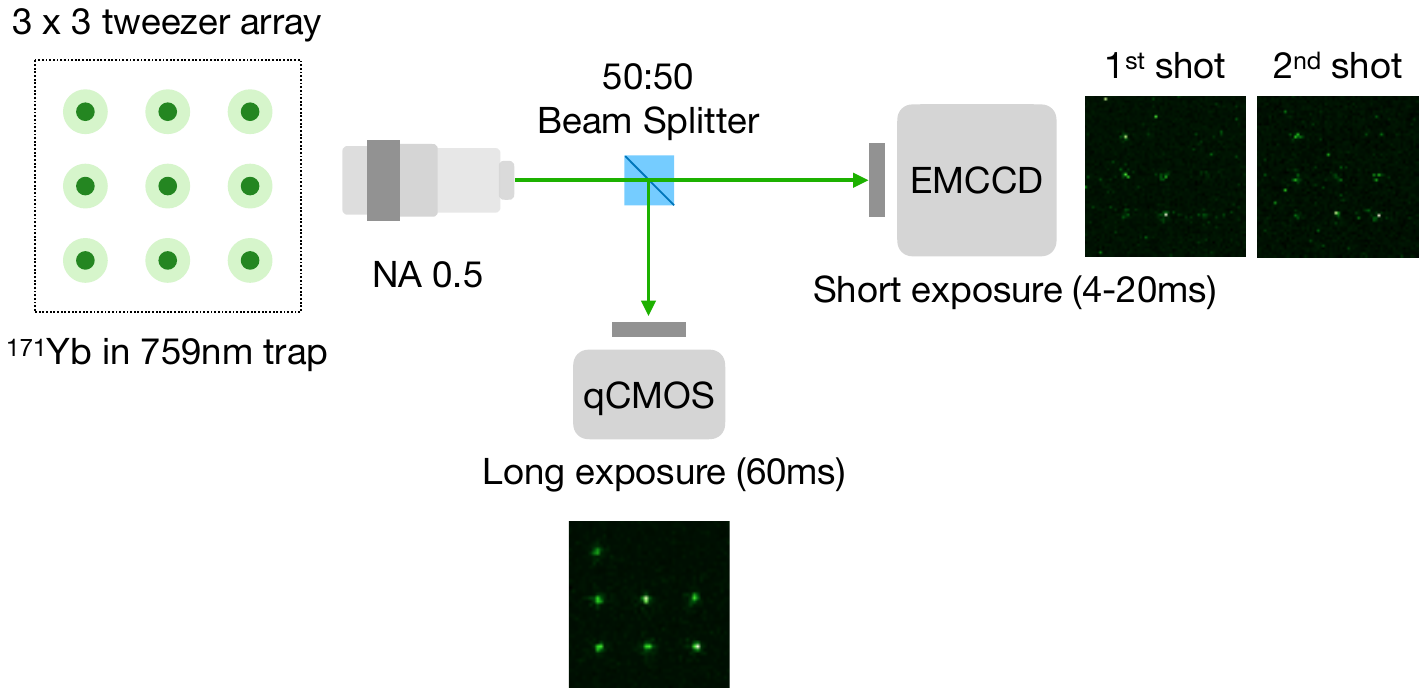}
\caption{\label{fig:setup}
\textbf{Experimental setup}
Simultaneous acquisition of readout and reference fluorescence images of a
\(3\times3\) \({}^{171}\mathrm{Yb}\) tweezer array. Fluorescence from atoms
trapped in a 759~nm optical tweezer array is collected with 556~nm imaging beam and an NA \(=0.5\)
objective. The fluorescence is then split by a 50:50 beam splitter. One arm is directed to an electron-multiplying CCD (EMCCD)
camera that acquires readout images with exposure times of \(4\)--\(20~\mathrm{ms}\), while the other
arm is directed to a qCMOS camera that acquires reference images with a \(60~\mathrm{ms}\) exposure for
reference-label generation. Two consecutive readout images were acquired in the EMCCD arm for each experimental shot.
}
\end{figure}

Atomic fluorescence was collected by an objective lens with numerical aperture $0.5$ (Mitutoyo G Plan Apo $50\times$) and split equally by a 50:50 beam splitter [Fig.\ref{fig:setup}]. One output arm was directed to an electron-multiplying CCD (EMCCD) camera (Andor iXon Ultra 897) with camera gain 300 for readout-image acquisition, while the other was directed to a qCMOS camera (Hamamatsu C15550-22UP) for reference-image acquisition. Readout and reference images were acquired simultaneously. Reference images were acquired at a fixed exposure time of $60~\mathrm{ms}$ for label generation, while pairs of consecutive readout images were recorded for denoiser training at exposure times $t_{img} = 4, 5, 8, 10, 12, 14, 16,$ and $20~\mathrm{ms}$. The corresponding average detected photon counts per ROI were $4.45$, $6.44$, $8.39$, $10.4$, $12.33$, $14.36$, $16.67$, and $20.77$, respectively. The 60 ms reference exposure is short compared with the characteristic imaging-loss time of approximately 5.26 s.

Images were acquired with a size of $64 \times 64$ pixels, and each atomic site was analyzed using a ROI with $5 \times 5$ pixels. The measured nearest-neighbor atom spacing was approximately $15$ pixels, corresponding to a physical atom-to-atom separation of $4.8~\mu\mathrm{m}$. For each readout exposure time, the dataset contained approximately $1.4\times 10^4$ to $3.0\times 10^4$ frames, corresponding to about $6.9\times 10^3$ to $1.5\times 10^4$ paired acquisition sets, since two readout images were recorded per set (see Appendix~\ref{app:data_preparation}). 

\section{Data-driven discriminative readout}

Data-driven discriminative readout directly learns the boundary between occupied and empty sites from labeled experimental data, rather than classifying through a fully specified pixel-level generative likelihood. In our implementation, each atom-site ROI is represented by a set of physically interpretable scalar features, and a logistic-regression classifier combines these features to produce an atom-present probability. Specifically, we extract (i) pixel-value-based, (ii) denoising-derived pixel-value-based, (iii) background-fitted camera-noise-model-based, and (iv) point-spread-function (PSF)-based features from the measured images and use them as inputs to a logistic-regression classifier.

The feature set used in the final classifier was selected by systematically evaluating a broad set of candidate descriptors, retaining only the most informative descriptors. The proposed framework is not restricted to the four features considered here; in principle, any physically or statistically motivated descriptor of the ROI can be incorporated into the discriminative classifier. We therefore examined additional candidates, including the fraction of atomic signal concentrated near the ROI center, the similarity between the denoiser-learned noise model and the observed noise, and residual spatial structure after subtracting a fitted PSF model from the ROI. Feature-addition analysis showed that these additional descriptors produced only small changes beyond the selected four features (see Appendix~\ref{app:ablation_candidates} for their definitions and Table~\ref{tab:supp_extra_features} for the results). Nevertheless, the framework remains extensible, and future studies may identify more informative descriptors that further improve classification performance or provide additional physical interpretability.

In this section, we describe the selected features and the logistic-regression model. We then analyze the resulting fidelity improvements and quantify the contribution of each feature. 
We benchmark our approach against two representative readout strategies: a conventional method which detects site occupancy based on thresholding of detected photon counts~\cite{cooper18_alkalineearthtweezers}, and an EMCCD noise-based Bayesian readout model following Martinez-Dorantes \textit{et al.}~\cite{martinezdorantes17} (see Appendix~\ref{app:baseline}).

\subsection{Feature description and logistic regression}
We selected four features that capture complementary aspects of the measured
image. For each atom site \(s\), we define a \(5\times5\) ROI centered at the
calibrated site position \(c_s=(x_s,y_s)\),
$\mathcal{R}_s=\{p=(x,y): |x-x_s|\le 2,\ |y-y_s|\le 2\}$. 
Here \(p\) indexes camera pixels. The raw image in experimental shot \(t\) is
denoted by \(I_t^{\mathrm{raw}}(p)\), in analog-to-digital units (ADU), and the
output of the trained denoiser is denoted by \(I_t^{\mathrm{den}}(p)\).
Background images are obtained from separate frames acquired before loading atoms in traps. We denote the corresponding raw and denoised background frames by
\(I_b^{\mathrm{raw,bg}}(p)\) and \(I_b^{\mathrm{den,bg}}(p)\), where \(b\)
indexes background frames. The mean raw and denoised backgrounds are
$
B^{\mathrm{raw}}(p)=\left\langle I_b^{\mathrm{raw,bg}}(p)\right\rangle_b,
\qquad
B^{\mathrm{den}}(p)=\left\langle I_b^{\mathrm{den,bg}}(p)\right\rangle_b .
$
We construct the following four scalar features from each ROI.

\paragraph{\textbf{Photon count in raw image.}}
The first feature is the total background-subtracted raw signal within the
\(5\times5\) ROI,
\begin{equation}
x_1(s,t)
=
G\sum_{p\in\mathcal{R}_s}
\left[I_t^{\mathrm{raw}}(p)-B^{\mathrm{raw}}(p)\right],
\end{equation}
where \(G=\kappa/g\) converts ADU counts into detected photoelectrons after EM
gain correction. This feature estimates the total photon signal collected
from the atom image. We use the camera sensitivity
\(\kappa=4.15~\mathrm{e^-}/\mathrm{ADU}\) and EM gain \(g=300\), giving
\(G\simeq 0.01383\). \\

\begin{figure}[!ht]
\centering
\includegraphics[
  width=\linewidth,
  trim={1cm 2.5cm 16cm 3cm},
  clip
]{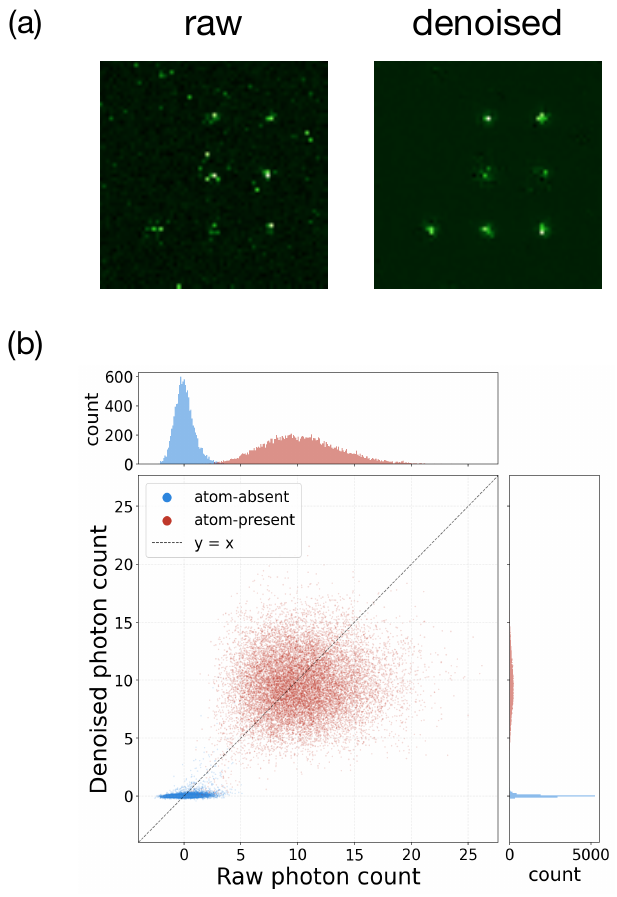}
\caption{\label{fig:denoise_squish}
\textbf{Denoising reduces photon-count variance in fast measurements.}
(a) Representative raw and denoised readout images at
\(t_{\mathrm{img}}=10~\mathrm{ms}\). Denoising suppresses camera/background
fluctuations while preserving the atom fluorescence spots.
(b) Raw-photon-count versus denoised-photon-count distribution at
\(t_{\mathrm{img}}=10~\mathrm{ms}\), with histograms for atom-absent and atom-present
ROIs. The denoised photon count preserves the photon scale of atom-present ROIs while
reducing the variance of the signal distributions for both atom-absent and atom-present ROIs, thereby
reducing overlap between the two classes.
}
\end{figure}

\paragraph{\textbf{Photon count in denoised image.}}
The second feature applies the same summation to the denoised image [Fig.\ref{fig:denoise_squish}(a)],
\begin{equation}
x_2(s,t)
=
G\sum_{p\in\mathcal{R}_s}
\left[I_t^{\mathrm{den}}(p)-B^{\mathrm{den}}(p)\right].
\end{equation}
This feature captures the atom-like signal after noise
suppression by the trained denoiser.
The denoising architecture is adapted from the Noise2NoiseFlow framework~\cite{maleky22, abdelhamed19}, which enables training without clean target images. We use two consecutively acquired images as paired noisy
observations for denoiser training. Specifically, we write the two images as
\(I_1=S+n_1\) and \(I_2=S+n_2\), where \(S\) denotes the underlying fluorescence
signal that is common to the pair, and \(n_1\) and \(n_2\) denote the
independent noise contributions in the two images. First, the images are denoised using a DnCNN module~\cite{zhang2017dncnn} which learns to remove noise by generating a clean-image estimate. The residual noise is then modeled by a normalizing-flow noise model~\cite{rezende2015normalizingflows, kingma2018glow} followed by DnCNN, which explicitly learns an invertible probability density function for the noise distribution. This allows the camera-noise statistics to be learned from data rather than imposed through a hand-crafted parametric model.

\begin{figure}[!ht]
\centering
\includegraphics[
  width=\linewidth,
  trim={0 0 0 0},
  clip
]{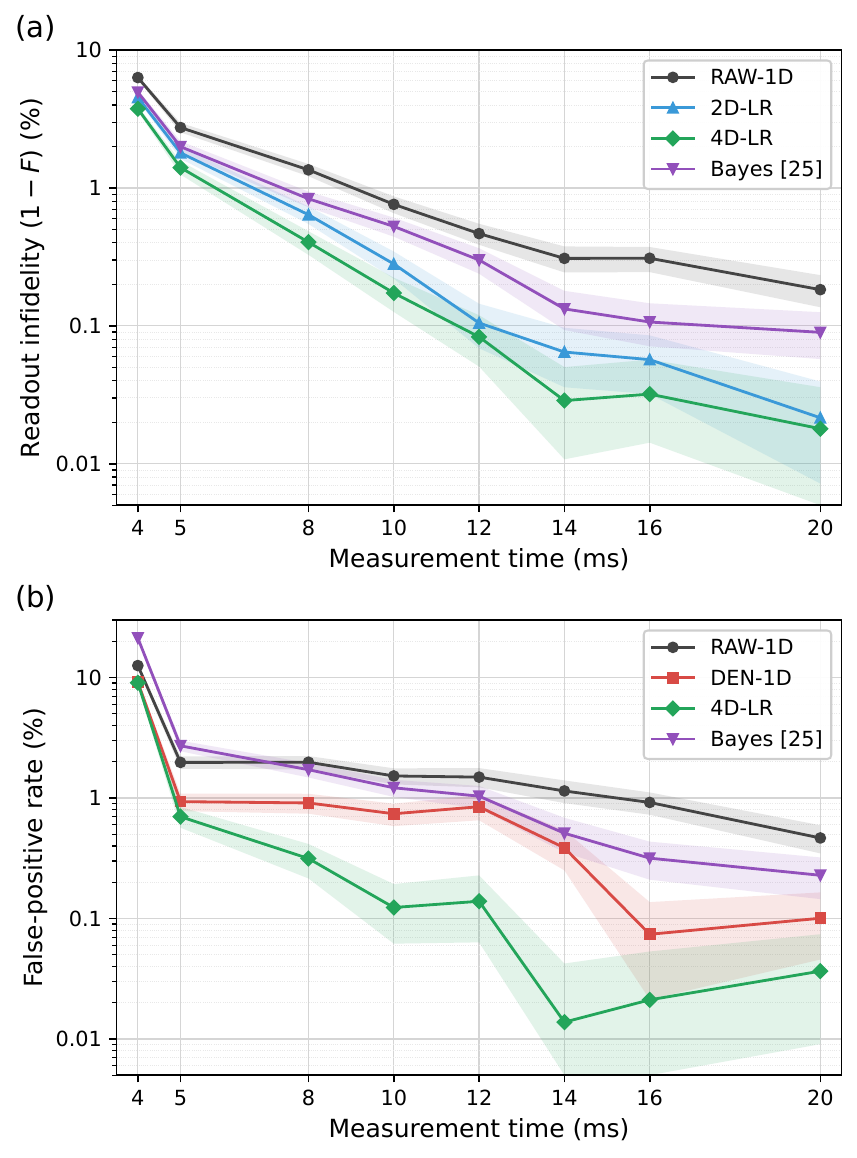}
\caption{\label{fig:fidelity}
\textbf{Readout fidelity and false-positive suppression.}
(a) Readout infidelity \(1-F\) as a function of measurement time for conventional thresholding on raw images (RAW-1D), two-feature logistic regression using two features (the raw ROI photon count and background-fitted camera-noise-model likelihood) (2D-LR), four-feature logistic regression (4D-LR), and the Bayesian baseline. The discriminative classifiers, 2D-LR and 4D-LR, provide the largest gains in the low-photon regime.
(b) False-positive rate (FPR) for the readout methods, including conventional thresholding on denoised images (DEN-1D). The denoised-ROI feature substantially lowers false-positive errors relative to RAW-1D and the Bayesian baseline.
Shading denotes pointwise 95\% confidence intervals from 5,000 frame-bootstrap
resamples of fixed out-of-fold predictions.
Bayes: Bayesian baseline adapted from Martinez-Dorantes \textit{et al.}~\cite{martinezdorantes17}.
}
\end{figure}

For our fluorescence data, we increased and optimized the depth of the DnCNN-based denoising module and simplified the flow-based noise model. Further details of the architecture and the optimized loss function are provided in Appendix~\ref{app:denoising}. As shown in Fig.\ref{fig:denoise_squish}(b), denoising reduces the variance of the signal distributions for both atom-present and atom-absent ROIs. This variance reduction suppresses false-positive events, as reflected in Fig.\ref{fig:fidelity}(b) and Table~\ref{tab:error_rates}.\\

\paragraph{\textbf{PSF template-matching score.}}
The third feature measures whether the spatial distribution of counts within the
background-subtracted ROI is consistent with an empirically measured
single-atom image. Here, the term point-spread-function (PSF) does not refer to
a fitted Gaussian or Airy model. Instead, we construct an empirical atom-image template directly from
atom-present images identified using the reference imaging arm. We then
define a normalized template-matching score, mathematically equivalent to the
Pearson correlation between the background-subtracted ROI and this empirical
template~\cite{Zhao2006NCC}. 
For each training fold, the empirical template \(T(p)\) is obtained by
averaging background-subtracted \(5\times5\) ROIs over all atom-present samples,
\begin{equation}
T(p)
=
\frac{1}{N_B}
\sum_{(s,t):y_{s,t}=1}
R_{s,t}(p)
\end{equation}
where \(N_B\) is the number of atom-present training samples and \(y_{s,t}\) denote the reference label where \(y_{s,t}=1\) indicating an atom-present sample and \(y_{s,t}=0\) indicating an atom-absent sample. After averaging,
negative values are clipped and the template is normalized to unit sum. The PSF feature is then defined as the Pearson correlation between the
background-subtracted ROI and the empirical template,
\begin{equation}
x_3(s,t)
=
\frac{
\sum_{p\in\mathcal{R}_s}
\left(R_{s,t}(p)-\overline{R}_{s,t}\right)
\left(T(p)-\overline{T}\right)
}{
\left\|R_{s,t}-\overline{R}_{s,t}\right\|_2
\left\|T-\overline{T}\right\|_2
}.
\end{equation}
The overbar denotes the mean and \(\|\cdot\|_2\) denotes the Euclidean norm over pixels over the \(5\times5\) ROI. This feature is designed to contribute spatial-shape
information. It measures whether the detected signal is
distributed like a single-atom fluorescence image.\\

\paragraph{\textbf{Background-fitted camera-noise log-likelihood.}}
This feature describes the similarity between the image's pixel distribution and EMCCD's noise distribution. Noise distribution is modeled by combining electron-multiplication statistics, readout noise, and
clock-induced-charge contributions~\cite{basden15, harpsoe12}.
For the likelihood feature, the camera model parameters are estimated from background images, yielding $\hat{\theta}_{\mathrm{bg}}$. We then define a background-fitted
EMCCD log-likelihood feature as
\begin{equation}
x_4(s,t)
=
\frac{1}{|\mathcal{R}_s|}
\sum_{p\in\mathcal{R}_s}
\log P_{\mathrm{EMCCD}}
\!\left(
I_t^{\mathrm{raw}}(p);\hat{\theta}_{\mathrm{bg}}
\right).
\end{equation}
Smaller values of \(x_4\) indicate that the observed ROI pixels are less
compatible with the EMCCD noise distribution fitted to background images and
are therefore more likely to contain atom fluorescence.
The detector-output likelihood feature is implemented here using an analytical
EMCCD output-noise model, because the detector response is well characterized
and this feature gave the best validation performance in our data. \\


\paragraph{\textbf{Logistic-regression Classifier}}
We use logistic regression as a simple discriminative model that assigns learned weights to the four features and maps their weighted sum to an atom-present probability.
Before fitting the logistic-regression classifier, each feature is standardized
using the mean and standard deviation computed from the training samples in the
corresponding cross-validation fold,
\begin{equation}
\widetilde{x}_k(s,t)
=
\frac{x_k(s,t)-\mu_k}{\sigma_k},
\qquad k=1,\ldots,4 .
\end{equation}
The standardized vector
$\widetilde{\mathbf{x}}(s,t)=(\widetilde{x}_1,\ldots,\widetilde{x}_4)$
defines a point in a four-dimensional feature space, where each dimension
corresponds to one scalar image descriptor. This vector is used as the input to
the logistic-regression classifier.
The atom-present probability is modeled as
\begin{equation}
P_{s,t}
=
P(y_{s,t}=1\mid s,t)
=
\mathrm{sigmoid}\!\left(
w_0+\sum_{k=1}^{4}w_k\widetilde{x}_k(s,t)
\right),
\end{equation}
where \(\sigmoid(u)=1/(1+e^{-u})\). Let \(w_0\) denote the bias term and
\(\mathbf{w}=(w_1,\ldots,w_4)\) the feature-weight vector. The weights are obtained by minimizing the
L2-regularized cross-entropy loss,

\begin{equation}
\hat{\mathbf{w}}
=
\arg\min_{\mathbf{w}}\mathcal{L}(\mathbf{w}) .
\end{equation}
\begin{equation}
\begin{aligned}
\mathcal{L}(\mathbf{w})
=
&-\sum_{(s,t)\in\mathcal{T}}
\ell_{s,t}
+
\frac{1}{2}\|\mathbf{w}_{1:4}\|_2^2 ,
\\
\ell_{s,t}
=
&\;y_{s,t}\log P_{s,t}
+
(1-y_{s,t})\log(1-P_{s,t}) .
\end{aligned}
\end{equation}
where \(\mathcal{T}\) denotes the training set.
The final binary prediction is obtained using a fixed threshold,
$\hat{y}_{s,t}=\mathbf{1}\!\left[P_{s,t}>0.5\right]$. All preprocessing steps, including the PSF-template construction, feature
standardization, and classifier fitting, are performed independently within
each training fold. Reported performance metrics are computed from out-of-fold
predictions.

\subsection{Fidelity improvement and error analysis}

We next quantify how the discriminative feature representation improves readout
performance relative to one-dimensional thresholding~\cite{cooper18_alkalineearthtweezers} and the Bayesian baseline~\cite{martinezdorantes17}.
Because the total fidelity alone can obscure the underlying error mechanism, we
also separately analyze the false-positive rate (FPR) and false-negative rate
(FNR) [Fig.\ref{fig:fidelity}]. Complete fidelity, FPR, and FNR values
across all measurement times are provided in Appendix~\ref{app:fidelity}.

The full four-feature classifier (4D-LR) combines complementary information
from photon counts, denoised intensity, camera-noise likelihood, and empirical
spatial structure to improve readout fidelity, with the largest absolute gains
in the fast imaging regime (\(t_{\mathrm{img}}\leq10~\mathrm{ms}\))
[Fig.\ref{fig:fidelity}(a)]. At \(10~\mathrm{ms}\), it achieves
\(99.83\%\) fidelity, comparable to the \(99.82\%\) obtained by RAW-1D
at \(20~\mathrm{ms}\), demonstrating an approximately twofold reduction
in the measurement time needed to reach this fidelity level. It also exceeds
\(99.9\%\) fidelity for all measurement times from \(12~\mathrm{ms}\)
onward.

The improvement also changes the error profile, substantially reducing
false-positive decisions [Fig.\ref{fig:fidelity}(b)]. At \(10~\mathrm{ms}\),
4D-LR reduces the FPR to \(0.12\%\), compared with \(1.52\%\) for RAW-1D
and \(1.22\%\) for the Bayesian baseline. Relative to the Bayesian baseline,
this suppression comes with a higher FNR, illustrating why total fidelity
alone does not fully characterize readout performance. 

Comparisons with simpler readout methods help explain the gain. A two-feature
classifier (2D-LR), using the raw photon count and camera-noise log-likelihood,
already improves on RAW-1D and the Bayesian baseline, while thresholding
denoised images (DEN-1D) primarily suppresses false positives. 
The detailed error-rate
comparisons are provided in Appendix~\ref{app:fidelity}.

These results highlight the practical advantage of learning the decision boundary directly from experimental data. Although a Bayesian decision rule minimizes classification error when the class-conditional distributions and priors are correctly specified, accurately modeling these distributions can be challenging in experiments. Our discriminative approach achieves higher empirical fidelity than the implemented Bayesian baseline without requiring a complete generative description of the observed images.

To quantify the contribution of each feature to 4D-LR, we remove one feature
at a time from the full classifier (Table~\ref{tab:ablation}). At 8 ms, removing the
camera-noise log-likelihood produces the largest fidelity degradation, followed by the raw photon count. Removing the denoised
photon count causes the largest increase in FPR, highlighting its role in
suppressing false-positive decisions. The PSF-template feature provides a
smaller but measurable fidelity gain. These results support combining
complementary features to improve both readout fidelity and the error profile.
Results at each measurement time and the feature-buildup analysis are provided
in Appendix~\ref{app:ablation}.

\begin{table}[!ht]
\centering
\caption{\label{tab:ablation}
\textbf{Leave-one-feature-out ablation analysis of the 4D-LR classifier.}
Ablation results are evaluated at \(t_{\mathrm{img}}=8~\mathrm{ms}\). Each feature is removed from the full model one at a time.
Positive \(\Delta F\) indicates fidelity degradation, and positive
\(\Delta\mathrm{FPR}\) indicates an increase in false-positive rate.
Changes are reported in percentage points (pp).
The largest value in each column is shown in bold.
}
\begin{tabular}{lcc}
\hline
Removed feature & $\Delta F$ (pp) & $\Delta \mathrm{FPR}$ (pp) \\
\hline
Raw Photon Count & 0.070 & 0.026 \\
Denoised Photon Count & 0.062 & \textbf{0.060} \\
PSF-template correlation & 0.059 & 0.009 \\
Camera-noise log-likelihood & \textbf{0.140} & 0.051 \\
\hline
\end{tabular}
\end{table}

\begin{figure*}[!ht]
\centering
\includegraphics[width=\textwidth]{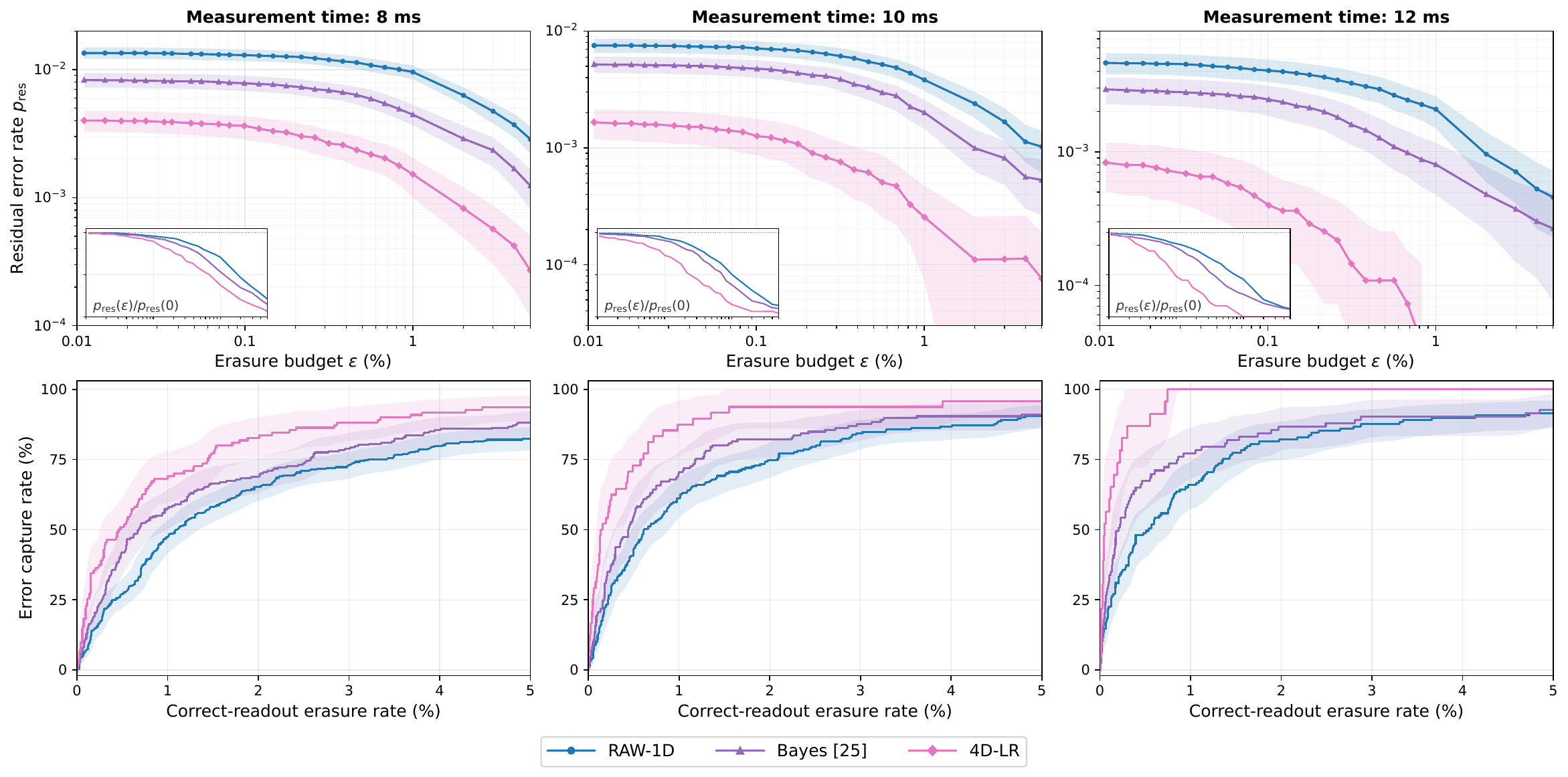}
\caption{\label{fig:erasure} 
\textbf{Confidence-aware measurement erasure conversion.}
Upper panels: residual unflagged error rate \(p_{\mathrm{res}}(\epsilon)\) as a function
of the erasure budget \(\epsilon\), shown for representative measurement  
times. ROIs are ranked by increasing decision confidence, and the
lowest-confidence fraction \(\epsilon\) is flagged as erasures. Insets show the
normalized residual error rate \(p_{\mathrm{res}}(\epsilon)/p_{\mathrm{res}}(0)\).
Lower panels: error-detection ROC curves obtained by varying the confidence
cutoff. The horizontal axis is the percentage of correct readouts (TP+TN)
flagged as erasures, and the vertical axis is the percentage of readout errors
(FP+FN) flagged as erasures. Each denominator is the corresponding total
before erasure filtering. Curves closer to the upper-left corner capture more
errors while flagging fewer correct outcomes.
Shaded bands denote pointwise 95\% confidence intervals obtained from 3,000
paired bootstrap resamples of acquisition frames, retaining all nine ROI
outcomes within each frame and keeping the out-of-fold predictions fixed.
}
\end{figure*}

\section{Confidence-Aware Measurement erasure conversion}

Conventional atom readout typically converts each ROI into a binary decision, even when the measured signal lies close to the decision boundary. However, for downstream applications such as erasure-aware error correction or atom-array assembly, it can be more useful to explicitly flag an ambiguous measurement as uncertain than to force it into an incorrect binary assignment. We therefore evaluate whether the scalar readout output value can also be used to identify decisions that are likely to be wrong.

Let \(s\) denote the scalar readout score, where
\(s=p_{\mathrm{atom}}\) for probabilistic classifiers and
\(s=\sigmoid[(x-x_{\mathrm{th}})/\sigma_{\mathrm{train}}]\) for the thresholding methods,
where \(x\) is the ROI sum, and the threshold \(x_{\mathrm{th}}\) and scale
\(\sigma_{\mathrm{train}}\) are determined from the training fold. We define the
confidence as the distance from the corresponding decision threshold,
$c = |s-s_{\mathrm{th}}|$,
where \(s_{\mathrm{th}}=0.5\).
Thus, near-threshold decisions are treated as more uncertain and can be
flagged as erasures.
This is analogous to using \textit{soft} measurement information in quantum error correction, where the magnitude of a likelihood or posterior score quantifies the reliability of a measured bit~\cite{ali2024softinfo}. It is also connected to erasure-conversion strategies, in which leakage-detected events are converted into located erasures rather than left as unflagged errors~\cite{wu22_erasure, chow2024leakageerasure}.

Let \(\mathcal{D}\) denote the set of all ROI decisions at a given exposure time before erasure filtering.
We rank these decisions by increasing confidence and flag
the lowest-confidence decisions as erasures. The erasure budget
\(\epsilon = |\mathcal{E}_{\epsilon}|/|\mathcal{D}|\) is the fraction of all
decisions flagged, where \(\mathcal{E}_{\epsilon}\) is the set of flagged decisions. Let
$
\mathcal{E}_{\epsilon}^{c}
\equiv
\mathcal{D}\setminus\mathcal{E}_{\epsilon}
$
denote the corresponding set of non-erased ROI decisions and \(\mathcal{M}\subset\mathcal{D}\) denote the set of readout errors. At each erasure budget \(\epsilon\), the residual error rate is the fraction
of non-erased decisions that are incorrect,
\begin{equation}
p_{\mathrm{res}}(\epsilon)
=
\frac{
\left|\mathcal{M}\cap\mathcal{E}_{\epsilon}^{c}\right|
}{
\left|\mathcal{E}_{\epsilon}^{c}\right|
}.
\end{equation}

The upper panels of Fig.\ref{fig:erasure} show how the residual error rate
\(p_{\mathrm{res}}\) varies with the erasure budget \(\epsilon\).

The lower panels of Fig.\ref{fig:erasure} compare error detection at a matched
cost of flagging correct readouts. With \(\mathcal{C}=\mathcal{D}\setminus\mathcal{M}\)
denoting the set of correct readouts, the horizontal and vertical axes give
the fractions of correct and incorrect readouts flagged as erasures, respectively,
\begin{align}
r_{\mathrm{corr}}(\epsilon)
&=\frac{|\mathcal{E}_{\epsilon}\cap\mathcal{C}|}{|\mathcal{C}|},\\
r_{\mathrm{err}}(\epsilon)
&=\frac{|\mathcal{E}_{\epsilon}\cap\mathcal{M}|}{|\mathcal{M}|}.
\end{align}
Higher error capture at a given correct-readout
erasure rate indicates better separation of errors from correct decisions.

The value of the readout score lies not only in the accuracy of its hard
decisions, but also in its ability to identify which decisions are
likely to be wrong. Figure \ref{fig:erasure} shows that 4D-LR provides
both advantages. It leaves the lowest residual unflagged error rate
over the tested erasure budgets, while its more rapidly decreasing
normalized curves indicate that low-confidence outcomes capture a
larger fraction of its initial errors. The benefit therefore extends
beyond higher readout fidelity to more effective identification of
unreliable decisions. This allows a limited erasure budget to be used
more effectively, reducing the errors that remain unflagged.

\section{Improved Atom Rearrangement via Measurement Erasure}

Atom rearrangement provides a direct application of confidence-aware
readout, because successful assembly depends on the reliability of the
sites selected as atom sources. A false-positive decision can cause
an empty site to be selected, leaving a defect in the target array,
whereas a false-negative decision mainly reduces the available atom
pool. This asymmetry motivates using measurement confidence to exclude
uncertain occupied-site predictions before rearrangement. Such
erasure-aware source selection can reduce incorrect source assignments,
provided that enough atoms remain to fill the target.

We quantified this effect using Monte Carlo simulations.
Atoms were initialized on a \(30\times30\) reservoir array with loading
probability \(p_{\mathrm{load}}=0.5\), and the rearrangement algorithm was applied to fill a square target array. The measured false-positive and false-negative rates were used as simulation
inputs and imaging-induced atom loss was included using
$P_{\mathrm{surv}}(t_{\mathrm{img}})
=
\exp(-\gamma_{\mathrm{loss}}t_{\mathrm{img}})$
with \(\gamma_{\mathrm{loss}}=1.9\times10^{-4}\,\mathrm{ms}^{-1}\).

We evaluated both single-round and multi-round rearrangement, defining the
success rate as the fraction of Monte Carlo trials that achieve defect-free
assembly under each protocol. In the single-round protocol, success requires
filling a fixed \(L\times L\) target after one readout-and-move cycle,
without an additional readout or refill within that round.
In the multi-round protocol, up to 20 cycles are allowed, with the target size
adjusted in each round according to the number of detected candidate atoms.
After each failed round, only atoms remaining in the attempted target region
are retained for the next round; unused reservoir atoms are discarded and
not reused. Success requires obtaining a defect-free array of at least
\(L\times L\) within the allowed number of rounds, so \(L\) specifies the
minimum target size in this protocol. For the erasure-aware methods,
low-confidence sites classified as occupied are excluded from the source pool
before each move.
The rearrangement algorithm and simulation procedure are described in
Appendix~\ref{app:rearrange}.

\begin{figure*}[!ht]
\centering
\includegraphics[width=\textwidth]{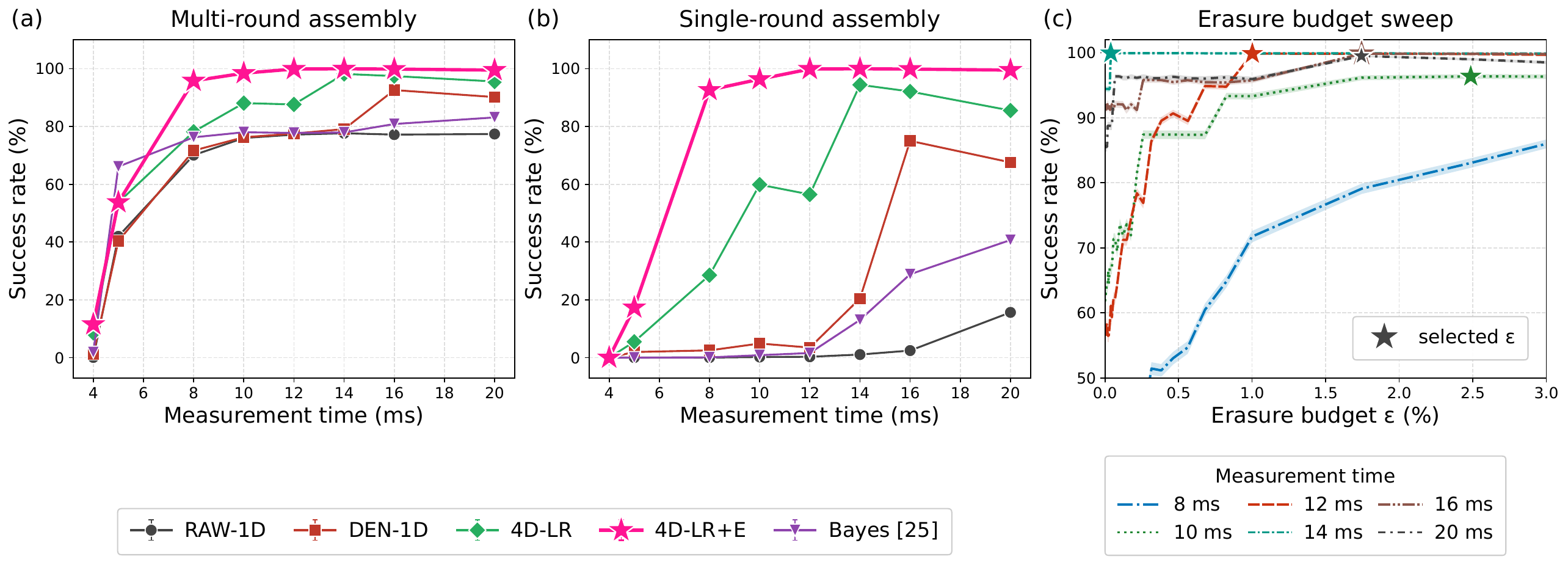}
\caption{\label{fig:rearrangement}
\textbf{Confidence-aware readout and measurement erasures improve atom rearrangement.}
(a) Multi-round success rate as a function of measurement time
\(t_{\mathrm{img}}\), with a minimum target size of \(20\times20\).
Simulations start from a \(30\times30\) reservoir array with loading
probability \(p_{\mathrm{load}}=0.5\).
4D-LR with erasure-aware source selection achieves success rates
above \(90\%\) for \(t_{\mathrm{img}}\geq8~\mathrm{ms}\).
(b) Single-round success rate as a function of measurement time
for a fixed \(20\times20\) target. The 4D-LR-based methods are the
only methods that yield substantial single-round success at
\(8\)--\(12~\mathrm{ms}\). At 10 ms, the success rate rises from
\(59.87\%\) without erasure filtering to \(96.37\%\) with it.
(c) Single-round success rate of 4D-LR as a function of erasure budget
for a fixed \(20\times20\) target. For each exposure, the star marks the
smallest erasure budget whose success rate is within one Monte Carlo standard
error of the maximum found in the sweep over budgets up to 25\%.
The displayed window shows \(8\)--\(20~\mathrm{ms}\), erasure budgets of
0--3\%, and success rates of at least 50\%. Stars are shown only when the
selected points fall within this window.
Shorter readout exposures
with confidence-based source selection can outperform longer readout exposures
without source filtering: single-round 4D-LR achieves \(96.37\%\) at 10 ms
with erasure filtering, versus \(85.53\%\) at 20 ms without it.
}
\end{figure*}

Confidence-aware source selection improves defect-free assembly in both
multi-round and single-round rearrangement. For a minimum target size of
\(20\times20\), 4D-LR with erasure-aware selection achieves multi-round
success rates above \(90\%\) for \(t_{\mathrm{img}}\geq8~\mathrm{ms}\)
[Fig.\ref{fig:rearrangement}(a)]. At 10 ms, erasure-aware selection increases
the predicted multi-round success rate from \(88.09\%\) to \(98.44\%\).
Its advantage is particularly clear
when only one rearrangement cycle is allowed. The 4D-LR-based methods
are the only methods that yield substantial success at
\(8\)--\(12~\mathrm{ms}\)
[Fig.\ref{fig:rearrangement}(b)], where incorrect source assignments
cannot be corrected by subsequent readout and rearrangement. At 10 ms,
erasure-aware selection increases the predicted single-round success rate
from \(59.87\%\) to \(96.37\%\), compared with \(4.90\%\) for DEN-1D,
\(0.86\%\) for the Bayesian baseline, and \(0.22\%\) for RAW-1D.

The dependence on erasure budget reveals a balance between source reliability
and atom availability. Rejecting uncertain candidates reduces false
source assignments, but excessive rejection leaves too few atoms to
fill the target [Fig.\ref{fig:rearrangement}(c)]. At an appropriate
erasure budget, shorter readout exposures with confidence-based selection can
outperform longer readout exposures without source filtering. For example,
single-round 4D-LR at 10 ms with erasure filtering achieves \(96.37\%\)
success, compared with \(85.53\%\) at 20 ms without erasure filtering.

\section{Discussion}
Our Bayesian baseline uses a physics-based likelihood model that explicitly
accounts for photon counting, camera noise, and atom loss. Its performance
depends on how accurately this model describes the fluorescence images for
occupied and empty sites.

With correctly specified class-conditional likelihoods and priors, assigning
each image to the class with the higher posterior probability minimizes the
expected classification error. The higher readout fidelity of 4D-LR
and its more effective identification of readout
errors suggest room to refine the likelihood model
used in our Bayesian baseline. The features used by 4D-LR suggest concrete
tests of the Bayesian likelihood model. For example, comparing simulated and
measured images of empty sites could reveal whether the model reproduces the
frequency of background fluctuations that resemble an atomic fluorescence spot.

One possible extension is therefore to retain the physical likelihood model as
the readout backbone while using calibration images to adjust the final
occupied/empty decision rule. Such a hybrid strategy~\cite{raina03} would preserve the interpretability
and data efficiency of the Bayesian model while correcting systematic decision
biases caused by experimental effects not fully captured in the analytical
likelihood model.

The feature-based structure of discriminative classifier makes the framework naturally extensible. More systematic feature discovery may reveal additional
informative observables, and such observables could in turn guide improved
generative modeling of the Bayesian likelihood. A particularly promising
direction is to use temporal information, such as photon-arrival statistics or
shot-to-shot intensity dynamics, which may provide information beyond the spatial ROI features considered here.

Several aspects are compatible with future experimental deployment. First, the denoising architecture does not rely on a specific model of the measurement device. It learns the camera-noise distribution from paired noisy images and can therefore be adapted to other fluorescence-imaging systems once appropriate training data are available. Such a variant would allow the logistic-regression classifier to operate without an explicitly defined camera-output model, which opens a route toward a camera model-agnostic readout pipeline.
Second, inference latency can be reduced in several ways. Because the dominant computational component is the convolutional denoiser, the pipeline is well suited to hardware acceleration. Recent work has demonstrated FPGA-accelerated convolutional-neural-network (CNN) inference for fluorescence-based qubit readout~\cite{su26_fpga_exposure_cnn}. Neural processing units (NPUs) with tensor-oriented architectures~\cite{kim24_tcp} can also efficiently execute neural-network inference workloads. Further reductions in memory traffic and
latency may be obtained using standard neural-network compression techniques
such as integer quantization and pruning~\cite{jacob18_integer_quantization,
han16_deep_compression,esser20_lsq}.

Our inference method reduces the number of detected photons needed to reach a given readout fidelity. Imaging could be made faster still by collecting those photons at a higher rate. For example, in our current setup, a 50:50 beam splitter diverts half of the collected fluorescence to a qCMOS camera for simultaneous reference imaging. Once the classifier has been trained and validated, the splitter could be removed for routine readout, approximately doubling the fluorescence photon flux reaching the EMCCD. This would yield a similar photon count in roughly half the exposure time and could allow comparable readout fidelity. Further improvements, such as a higher numerical aperture, a higher scattering rate, or more efficient cooling, could help achieve faster imaging with high atom survival.

In quantum error correction, shorter readout times can reduce idle errors
on data qubits during measurement~\cite{veroni24, graham23_midcircuit},
while confidence-based erasure conversion can aid decoding by flagging
unreliable measurement outcomes as erasures~\cite{wu22_erasure}.
Together, these effects can lower the logical error rate. A systematic
study of the tradeoff between imaging time, measurement fidelity, erasure
budget, and the number of measurement rounds remains an important
direction for future work.

\section{Conclusion}

We demonstrated confidence-aware fluorescence readout of individual \(^{171}\mathrm{Yb}\) atoms in optical tweezers using a feature-based discriminative classifier. By combining raw and denoised photon counts, a PSF-template correlation score, and a background-fitted camera-noise log-likelihood, the logistic-regression model learns the occupied/empty decision boundary directly from labeled experimental data. This formulation retains physically motivated information while learning the decision rule without requiring complete class-conditional likelihood models for the fluorescence images.

In the evaluated dataset, the four-feature classifier achieved higher occupancy-readout fidelity than conventional photon-count thresholding and the implemented Bayesian baseline. At an imaging exposure of 10~ms, it reached 99.83\% fidelity, comparable to the 99.82\% obtained by conventional thresholding at 20~ms. At this operating point, the classifier thus achieved comparable fidelity at half the exposure used by the threshold baseline. The improvement was accompanied by substantial suppression of false-positive decisions.

Beyond improved binary classification, the classifier score also provides useful information about the reliability of individual readout outcomes. Flagging low-confidence outcomes as measurement-level erasures leaves fewer residual unflagged errors than the threshold and Bayesian baselines at matched erasure budgets. For atom rearrangement, this information enables uncertain occupied-site predictions to be excluded from the source pool. Monte Carlo simulations parameterized by the measured error rates predict improved defect-free assembly through this confidence-based source selection. These results illustrate how readout confidence and the unequal consequences of false-positive and false-negative decisions can be used in downstream operations such as atom rearrangement.

The feature-based formulation is naturally extensible. Additional physically or statistically motivated descriptors can be incorporated and evaluated for a given imaging system or downstream task. This flexibility allows the readout model to be refined as new experimentally informative features are identified. Our framework thus provides a practical basis for combining accurate occupancy discrimination with confidence-guided atom selection, and for investigating its use in larger arrays and future mid-circuit measurement protocols.

\begin{acknowledgments}
 We thank Kangjin Kim, JuYoung Park, Sunhwa Hwang, and Seungtaek Oh for helpful discussions. This work was supported by the National Research Foundation of Korea(NRF) under Grant No. RS-2025-25464182, RS-2023-NR119928, and RS-2023-00283259.
\end{acknowledgments}

\clearpage
\onecolumngrid
\appendix
\renewcommand{\thesection}{\Alph{section}}

\titleformat{\section}[block]
  {\normalfont\normalsize\bfseries\centering}
  {APPENDIX~\thesection:}
  {0pt}
  {\\\MakeUppercase}

\renewcommand{\thefigure}{\Alph{section}\arabic{figure}}
\renewcommand{\thetable}{\Alph{section}\arabic{table}}
\renewcommand{\theequation}{\Alph{section}\arabic{equation}}

\makeatletter
\@addtoreset{equation}{section}
\@addtoreset{figure}{section}
\@addtoreset{table}{section}
\makeatother

\setcounter{figure}{0}
\setcounter{table}{0}
\setcounter{equation}{0}

\section{Data preparation and cross-validation protocol}
\label{app:data_preparation}

For supervised evaluation, we used the qCMOS reference image to label each site as occupied or empty. The EMCCD readout and qCMOS reference images were acquired simultaneously, with exposure times of 4--20~ms and 60~ms, respectively. An atom that fluoresces early in the exposure and is then lost or stops scattering can produce enough signal to be detected in the short readout image, yet leave too little integrated signal to cross the threshold for an occupied site in the reference image. We flagged frames containing at least one site labeled empty in the qCMOS reference for which the background-subtracted ROI signal in the EMCCD image exceeded threshold. We visually inspected the flagged frames and retained frames identified as EMCCD false positives and excluded the remaining flagged frames in their entirety. We additionally excluded acquisition sets in which no atoms were detected at any site of the \(3\times3\) array, because fully empty images were empirically found to degrade denoising-model training.

Table~\ref{tab:data_split} summarizes the number of acquired frames, the number
of test frames remaining after filtering, and the atom-loading fraction in the
filtered test set. The loading fraction was computed as the number of
atom-present sites divided by the total number of analyzed sites in the
filtered test set. These loading fractions were also used when reporting the
prior-weighted readout fidelity defined in the main text.

We evaluated all methods on the same filtered frames using five-fold cross-validation.
In each split, we used four folds to fit the readout method and its preprocessing
parameters, then applied them to the fifth fold. The Noise2NoiseFlow denoiser
was fixed across all folds. We computed all classification metrics from the
held-out predictions.

\begin{table}[!ht]
\caption{
\label{tab:data_split}
\textbf{Data split, filtering, and loading fraction.}
Total frames denote the number of acquired EMCCD readout images for each
measurement time. The data were split into training, validation, and test sets
using a 6:2:2 ratio. Filtered test frames denote the number of test frames
remaining after excluding fully empty acquisition sets and samples with
label-inconsistent atom appearance between the readout and reference images.
The loading rate is computed from the filtered test set as the fraction of
atom-present sites among all analyzed sites. This loading rate varies because
each measurement-time condition was acquired in an independent experimental run.
}
\begin{ruledtabular}
\scriptsize
\setlength{\tabcolsep}{3pt}
\begin{tabular}{ccccc}
Exp. & Total & Test & Filtered & Loading \\
(ms) & frames & frames & test frames & rate (\%) \\
\hline
4  & 30,000 & 5,357 & 5,284 & 80.28 \\
5  & 13,770 & 2,458 & 2,444 & 37.61 \\
8  & 15,952 & 3,190 & 3,024 & 56.78 \\
10 & 15,986 & 3,197 & 3,084 & 59.09 \\
12 & 16,000 & 3,200 & 3,073 & 71.37 \\
14 & 16,000 & 3,200 & 3,099 & 73.99 \\
16 & 15,994 & 3,198 & 3,128 & 66.35 \\
20 & 15,996 & 3,199 & 3,098 & 60.72 \\
\end{tabular}
\end{ruledtabular}
\end{table}

\clearpage
\section{readout baselines and evaluation metrics}
\label{app:baseline}

\subsection{Threshold baseline}
As a conventional baseline, we use a thresholding method based on the ROI-integrated signal \(x = \sum_{(i,j)\in\mathrm{ROI}} c_{ij}\), where \(c_{ij}\) denotes the gain-calibrated pixel value at ROI coordinate \((i,j)\).
The signal distributions for atom-absent and atom-present ROIs are modeled from ground-truth-labeled samples, with the atom-absent distribution represented by a Gaussian peak and the full-exposure atom-present signal distribution \(P(x\mid B)\) represented by a Gaussian peak with separately estimated mean and variance. Atom loss during the imaging is incorporated by writing the signal distribution for initially occupied sites as a survival-loss mixture,
\begin{equation}
P(x\mid B_{\mathrm{init}})
=
s\,P(x\mid B)
+
(1-s)\,P(x\mid L),
\end{equation}
where \(P(x\mid L)\) is a exponential decay tail describing atoms that emit only a fraction of the full-exposure signal before being lost. The survival probability is approximated to first order as
\begin{equation}
s = 1 - r_{\mathrm{loss}} t_{\mathrm{exp}},
\qquad
r_{\mathrm{loss}} = 0.019\%\ \mathrm{ms}^{-1}.
\end{equation}
For each measurement time condition, the classification threshold \(x_{\mathrm{th}}\) is chosen to maximize the prior-weighted fidelity,
\begin{equation}
\begin{aligned}
\mathcal{F}(x_{\mathrm{th}})
&=
1-
\Bigl[
(1-f)\,P(x\ge x_{\mathrm{th}}\mid D) \\
&\qquad\quad
+
f\,P(x<x_{\mathrm{th}}\mid B_{\mathrm{init}})
\Bigr].
\end{aligned}
\end{equation}
where \(f\) is the empirical atom-loading rate derived from the training data.

\subsection{Bayesian baseline}

As a physics-based baseline method, we implemented a Bayesian readout model following Martinez-Dorantes \textit{et al.}~\cite{martinezdorantes17}. Given an observed signal value $c$ and occupancy class $S\in\{B,D\}$, the posterior is obtained from Bayes' theorem,
\begin{equation}
P^{\mathrm{post}}(S\mid c)
=
\frac{P(c\mid S)\,P^{\mathrm{pri}}(S)}
{\sum_{S'} P(c\mid S')\,P^{\mathrm{pri}}(S')}.
\end{equation}
In our implementation, the class-conditional likelihood $P(c\mid S)$ explicitly incorporates class-dependent photon-emission statistics including leakage, Poissonian clock-induced charge (CIC), stochastic EM-gain amplification, and Gaussian readout noise. In particular, the total electron-count distribution is modeled as a convolution of the class-dependent photon distribution with a Poisson CIC process, and the resulting camera response is modeled by an Erlang amplification distribution convolved with Gaussian readout noise. Atom loss during imaging is incorporated as a survival-loss mixture in the atom-present likelihood, using the measured loss rate \(\gamma_{\mathrm{loss}}\) to model the fluorescence duration shared by all ROI pixels. Pixel likelihoods are multiplied conditional on that duration before integrating over the common loss time. The camera-noise parameters remain fixed at their calibration values. The prior probabilities \(P^{\mathrm{pri}}(S)\) were estimated from the
training partition of each cross-validation split. Specifically,
\(P^{\mathrm{pri}}(B)\) was set to the atom-present fraction in the
training labels, and \(P^{\mathrm{pri}}(D)=1-P^{\mathrm{pri}}(B)\). These priors
were then held fixed when evaluating the Bayesian posterior on the
corresponding held-out test fold.

Unlike the original column-by-column formulation of Ref.~\cite{martinezdorantes17}, our geometry permits direct inference on a fixed two-dimensional ROI around each site, since the nearest-neighbor spacing is much larger than the imaging PSF width. We therefore perform pixel-by-pixel Bayesian inference within each ROI, using the mean atom-present image measured within each training fold as a PSF-like two-dimensional signal template. The resulting posterior log-probability is then thresholded to obtain the Bayesian discriminator. 

\subsection{Evaluation metrics}

Throughout this work, readout fidelity is evaluated directly from the confusion matrix against reference labels, rather than estimated from parametric fits to the observed count distributions. Let $f = N_{\mathrm{atom}}/N_{\mathrm{total}}$ denote the atom-loading fraction, $F_1 = \mathrm{TP}/(\mathrm{TP}+\mathrm{FN})$ the atom-present readout fidelity, and $F_0 = \mathrm{TN}/(\mathrm{TN}+\mathrm{FP})$ the atom-absent readout fidelity. We then define the total fidelity as
\begin{equation}
F = (1-f)F_0 + fF_1,
\end{equation} 
and the corresponding infidelity as
\begin{equation}
\epsilon = 1 - F.
\end{equation} where $f$ is a loading fraction.
All reported classification metrics were evaluated under cross-validation. In particular, unless otherwise noted, thresholds and supervised classifiers were determined on the training folds and then applied to test folds under 5-fold cross-validation. All nine sites in an acquisition frame belong to the same fold. Statistical uncertainties
are pointwise 95\% percentile intervals from bootstrap resampling of whole
frames with the out-of-fold predictions held fixed: 5,000 resamples for fidelity
and error rates, and 3,000 paired resamples for erasure curves. Here, \(\mathrm{FPR}=\mathrm{FP}/N_{\mathrm{empty}}\),
\(\mathrm{FNR}=\mathrm{FN}/N_{\mathrm{atom}}\), and
\(F=1-(\mathrm{FP}+\mathrm{FN})/N_{\mathrm{total}}\).

\clearpage
\section{Detailed readout metrics}
\label{app:fidelity}

For completeness, Tables~\ref{tab:fidelity_only} and~\ref{tab:error_rates}
report the readout fidelity and the corresponding false-positive and
false-negative rates, respectively, for all methods across the full range of
measurement times. These numerical results supplement the trends shown in
Fig.\ref{fig:fidelity} and support the comparisons discussed in the main text.

At the fixed decision thresholds used in this work, 2D-LR achieves higher readout fidelity than both
RAW-1D and the Bayesian baseline across all measurement times. The improvement
is largest in the fast readout regime (\(t_{\mathrm{img}}\leq10~\mathrm{ms}\)),
where the signal distributions for atom-present and atom-absent ROIs strongly overlap and residual
mismatch in the generative model is expected to have the strongest effect.

For DEN-1D, the same one-dimensional thresholding procedure used for RAW-1D is
applied after image denoising. DEN-1D reaches a fidelity comparable to the
Bayesian baseline, suggesting that image-domain noise suppression can recover
a large part of the gain obtained by explicit camera-noise modeling. Its error
profile is different, however, as denoising mainly suppresses false positives.
At \(4~\mathrm{ms}\), for example, the FPR decreases from \(12.56\%\)
with RAW-1D to \(9.14\%\) with DEN-1D.

The full four-dimensional feature set further improves the decision boundary.
The largest absolute fidelity gains occur in the fast imaging regime, where
classification errors are dominated by overlap between the atom-present and atom-absent signal distributions.
For \(t_{\mathrm{img}}>10~\mathrm{ms}\), the baseline fidelity is already close to unity
and the remaining room for improvement is small. Nevertheless, 4D-LR exceeds
\(99.9\%\) fidelity from \(12~\mathrm{ms}\) onward. Its \(99.83\%\)
fidelity at \(10~\mathrm{ms}\) is comparable to the \(99.82\%\) reached
by RAW-1D at \(20~\mathrm{ms}\), supporting the measurement-time reduction
discussed in the main text.

The Bayesian baseline tends to classify ambiguous sites as occupied,
suppressing false negatives at the cost of a relatively high FPR. This
behavior is particularly clear at \(4~\mathrm{ms}\), where its FNR is
\(0.94\%\) but its FPR is \(21.26\%\). By contrast, 4D-LR reduces the
FPR to \(9.06\%\) while keeping the FNR at \(2.44\%\). The corresponding
FPR reduction of \(12.20\) percentage points is accompanied by a smaller
FNR increase of \(1.50\) percentage points. Across the full measurement-time
range, 4D-LR consistently has a lower FPR and a higher FNR than the Bayesian
baseline (Table~\ref{tab:error_rates}).

\begin{table*}[!ht]
\caption{
\label{tab:fidelity_only}
\textbf{Readout fidelity.}
Entries are fidelity [95\% CI], in percent. Intervals use 5,000 whole-frame
bootstrap resamples of fixed out-of-fold predictions.
Bold values mark the highest fidelity at each measurement time.
RAW-1D: Conventional thresholding in raw image;
DEN-1D: Conventional thresholding in denoised image;
Bayes: Bayesian baseline adapted from Martinez-Dorantes \textit{et al.}~\cite{martinezdorantes17}.
2D-LR: two-dimensional logistic regression using raw-image features;
4D-LR: four-dimensional logistic regression;
}
\begin{ruledtabular}
\footnotesize
\setlength{\tabcolsep}{3pt}
\begin{tabular}{cc|ccccc}
Meas. & Mean & RAW-1D & DEN-1D & Bayes~\cite{martinezdorantes17} & 2D-LR & 4D-LR \\
(ms) & Photon & Fid. (\%) & Fid. (\%) & Fid. (\%) & Fid. (\%) & Fid. (\%) \\
\hline
4 & 4.45 & $93.68\,[93.46,\,93.90]$ & $94.97\,[94.76,\,95.16]$ & $95.05\,[94.85,\,95.25]$ & $95.47\,[95.29,\,95.66]$ & $\mathbf{96.25}\,[96.08,\,96.42]$ \\
5 & 6.44 & $97.26\,[97.04,\,97.47]$ & $97.99\,[97.81,\,98.17]$ & $98.01\,[97.82,\,98.19]$ & $98.20\,[98.02,\,98.37]$ & $\mathbf{98.60}\,[98.44,\,98.75]$ \\
8 & 8.39 & $98.65\,[98.51,\,98.78]$ & $99.16\,[99.04,\,99.27]$ & $99.17\,[99.06,\,99.28]$ & $99.36\,[99.27,\,99.45]$ & $\mathbf{99.60}\,[99.51,\,99.67]$ \\
10 & 10.40 & $99.24\,[99.13,\,99.34]$ & $99.58\,[99.51,\,99.66]$ & $99.47\,[99.39,\,99.56]$ & $99.72\,[99.65,\,99.78]$ & $\mathbf{99.83}\,[99.78,\,99.87]$ \\
12 & 12.33 & $99.53\,[99.45,\,99.61]$ & $99.73\,[99.67,\,99.79]$ & $99.70\,[99.63,\,99.76]$ & $99.90\,[99.86,\,99.93]$ & $\mathbf{99.92}\,[99.88,\,99.95]$ \\
14 & 14.36 & $99.69\,[99.62,\,99.76]$ & $99.89\,[99.85,\,99.93]$ & $99.87\,[99.82,\,99.91]$ & $99.94\,[99.90,\,99.96]$ & $\mathbf{99.97}\,[99.95,\,99.99]$ \\
16 & 16.67 & $99.69\,[99.63,\,99.75]$ & $99.96\,[99.94,\,99.99]$ & $99.89\,[99.85,\,99.93]$ & $99.94\,[99.91,\,99.97]$ & $\mathbf{99.97}\,[99.94,\,99.99]$ \\
20 & 20.77 & $99.82\,[99.77,\,99.86]$ & $99.96\,[99.94,\,99.98]$ & $99.91\,[99.87,\,99.94]$ & $99.98\,[99.96,\,99.99]$ & $\mathbf{99.98}\,[99.96,\,100.00]$ \\
\end{tabular}
\end{ruledtabular}
\end{table*}

\begin{table*}[!ht]
\caption{
\label{tab:error_rates}
\textbf{False-positive and false-negative rates.}
FPR and FNR values are reported in percentage.
RAW-1D: Conventional thresholding in raw image;
DEN-1D: Conventional thresholding in denoised image;
Bayes: Bayesian baseline adapted from Martinez-Dorantes \textit{et al.}~\cite{martinezdorantes17}.
2D-LR: two-dimensional logistic regression using raw-image features;
4D-LR: four-dimensional logistic regression;
}
\begin{ruledtabular}
\scriptsize
\begin{tabular}{cc|cc|cc|cc|cc|cc}
Meas. & Mean
& \multicolumn{2}{c|}{RAW-1D}
& \multicolumn{2}{c|}{DEN-1D}
& \multicolumn{2}{c|}{Bayes~\cite{martinezdorantes17}}
& \multicolumn{2}{c|}{2D-LR}
& \multicolumn{2}{c}{4D-LR} \\
(ms) & Photon
& FPR (\%) & FNR (\%)
& FPR (\%) & FNR (\%)
& FPR (\%) & FNR (\%)
& FPR (\%) & FNR (\%)
& FPR (\%) & FNR (\%) \\
\hline
4 & 4.45 & 12.56 & 4.78 & 9.14 & 4.02 & 21.26 & 0.94 & 11.62 & 2.78 & 9.06 & 2.44 \\
5 & 6.44 & 1.97 & 4.01 & 0.93 & 3.80 & 2.71 & 0.80 & 1.01 & 3.11 & 0.70 & 2.57 \\
8 & 8.39 & 1.98 & 0.87 & 0.91 & 0.79 & 1.72 & 0.16 & 0.54 & 0.72 & 0.31 & 0.47 \\
10 & 10.40 & 1.52 & 0.23 & 0.74 & 0.20 & 1.22 & 0.05 & 0.25 & 0.30 & 0.12 & 0.21 \\
12 & 12.33 & 1.49 & 0.06 & 0.85 & 0.04 & 1.04 & 0.01 & 0.13 & 0.10 & 0.14 & 0.06 \\
14 & 14.36 & 1.14 & 0.01 & 0.39 & 0.01 & 0.51 & 0.00 & 0.07 & 0.06 & 0.01 & 0.03 \\
16 & 16.67 & 0.92 & 0.00 & 0.07 & 0.02 & 0.32 & 0.00 & 0.05 & 0.06 & 0.02 & 0.04 \\
20 & 20.77 & 0.47 & 0.00 & 0.10 & 0.00 & 0.23 & 0.00 & 0.04 & 0.01 & 0.04 & 0.01 \\
\end{tabular}
\end{ruledtabular}
\end{table*}

\clearpage
\section{4D-LR ablation study}
\label{app:ablation}

\subsection{Ablation of the selected four features}

To identify the contribution of each feature in the 4D-LR classifier, we
performed two complementary ablation analyses. First, we performed a leave-one-feature-out study, in which
each feature was removed from the full 4D-LR classifier. This complementary
analysis quantifies how much the final classifier depends on each feature after
all feature correlations have been taken into account. Second, we carried out a
feature-buildup study, starting from the raw photon count feature and sequentially
adding the background-fitted EMCCD log-likelihood, the denoised photon count 
feature, and the PSF template-matching score. This analysis shows how each
additional feature improves the classifier when combined with the features
already included.

Table~\ref{tab:ablation} in the main text summarizes the reported changes in fidelity and FPR
at \(t_{\mathrm{img}}=8~\mathrm{ms}\), and Table~\ref{tab:supp_dropout_4d} provides
the results at each measurement time, including changes in FNR.
The leave-one-feature-out analysis quantifies the contribution of each
physics- or AI-derived feature to the full classifier. The camera-noise
log-likelihood generally has the largest effect on fidelity for
\(t_{\mathrm{img}}\leq10~\mathrm{ms}\), while denoised intensity has a clear role in false-positive suppression.
Removing the PSF template also causes a measurable fidelity loss, although
its effect is generally smaller than that of the camera-noise likelihood.
Because feature-buildup results depend on the order in which features are
added, we treat that analysis as an interpretability diagnostic.

Table~\ref{tab:supp_dropout_4d}
confirms that no single feature alone explains the full performance gain:
removing the EMCCD log-likelihood generally causes
the largest degradation for \(t_{\mathrm{img}}\leq10~\mathrm{ms}\). These results support the
interpretation that 4D-LR improves readout by combining complementary
information from photon number, camera-output likelihood, denoised intensity,
and empirical spatial structure.
Table~\ref{tab:supp_buildup_4d_rawstart} shows that the EMCCD log-likelihood
provides the largest initial improvement beyond raw-count-only logistic regression,
especially in the low-photon regime. The denoised photon count feature provides an
additional reduction in false-positive rate, while the PSF template-matching
score gives a smaller final improvement. 

\begin{table*}[!ht]
\caption{
\label{tab:supp_dropout_4d}
\textbf{Leave-one-feature-out ablation of the 4D-LR classifier.}
Each entry reports the performance degradation after removing one feature from
the full 4D-LR model. \(\Delta F=F_{\mathrm{full}}-F_{\mathrm{drop}}\),
\(\Delta\mathrm{FPR}=\mathrm{FPR}_{\mathrm{drop}}-\mathrm{FPR}_{\mathrm{full}}\),
and \(\Delta\mathrm{FNR}=\mathrm{FNR}_{\mathrm{drop}}-\mathrm{FNR}_{\mathrm{full}}\).
All values are reported in percentage points.
}
\begin{ruledtabular}
\scriptsize
\begin{tabular}{c|ccc|ccc|ccc|ccc}
Exp.
& \multicolumn{3}{c|}{Drop raw amp}
& \multicolumn{3}{c|}{Drop denoised amp}
& \multicolumn{3}{c|}{Drop EMCCD likelihood}
& \multicolumn{3}{c}{Drop PSF template} \\
(ms)
& \(\Delta F\) & \(\Delta\)FPR & \(\Delta\)FNR
& \(\Delta F\) & \(\Delta\)FPR & \(\Delta\)FNR
& \(\Delta F\) & \(\Delta\)FPR & \(\Delta\)FNR
& \(\Delta F\) & \(\Delta\)FPR & \(\Delta\)FNR \\
\hline
4 & 0.18 & 0.30 & 0.15 & 0.11 & 0.54 & 0.01 & 0.26 & 0.62 & 0.18 & 0.18 & 0.54 & 0.09 \\
5 & 0.11 & 0.03 & 0.24 & 0.05 & 0.17 & -0.16 & 0.25 & 0.06 & 0.56 & 0.17 & 0.05 & 0.37 \\
8 & 0.07 & 0.03 & 0.10 & 0.06 & 0.06 & 0.06 & 0.14 & 0.05 & 0.21 & 0.06 & 0.01 & 0.10 \\
10 & 0.01 & -0.02 & 0.02 & 0.05 & 0.04 & 0.06 & 0.04 & 0.01 & 0.07 & 0.02 & 0.00 & 0.04 \\
12 & 0.00 & 0.01 & 0.00 & 0.01 & 0.01 & 0.01 & 0.03 & 0.06 & 0.02 & 0.02 & 0.01 & 0.02 \\
14 & 0.00 & 0.00 & 0.00 & 0.02 & 0.06 & 0.00 & 0.01 & 0.00 & 0.01 & 0.00 & 0.00 & 0.00 \\
16 & 0.00 & 0.00 & -0.01 & 0.00 & 0.03 & -0.02 & 0.01 & 0.00 & 0.02 & 0.01 & 0.00 & 0.01 \\
20 & 0.00 & 0.00 & 0.01 & 0.01 & 0.02 & 0.00 & 0.01 & 0.00 & 0.02 & 0.00 & -0.01 & 0.01 \\
\end{tabular}
\end{ruledtabular}
\end{table*}

\begin{table*}[!ht]
\caption{
\label{tab:supp_buildup_4d_rawstart}
\textbf{Feature-buildup ablation of the 4D-LR classifier.}
The classifier is built by starting from the raw photon count feature and
sequentially adding the background-fitted camera-noise log-likelihood, the denoised
photon count, and the PSF template-matching score. Values are reported in
percentage. Numbers in parentheses indicate the change relative to the previous
buildup step, in percentage points. For fidelity, positive changes indicate
improvement; for FPR and FNR, negative changes indicate error reduction. The
final step corresponds to the full 4D-LR classifier.
}
\begin{ruledtabular}
\scriptsize
\begin{tabular}{ccccc}
Exp. & Feature set & Fid. (\%) & FPR (\%) & FNR (\%) \\
(ms) &  &  &  &  \\
\hline
4 & Raw amp & 93.99 & 15.71 & 3.63 \\
4 & + EMCCD LL & 95.47 (+1.48) & 11.62 (-4.08) & 2.78 (-0.85) \\
4 & + Denoised amp & 96.07 (+0.60) & 9.61 (-2.02) & 2.53 (-0.25) \\
4 & + PSF template & 96.25 (+0.18) & 9.06 (-0.54) & 2.44 (-0.09) \\
\hline
5 & Raw amp & 97.14 & 1.67 & 4.84 \\
5 & + EMCCD LL & 98.20 (+1.06) & 1.01 (-0.66) & 3.11 (-1.73) \\
5 & + Denoised amp & 98.42 (+0.22) & 0.75 (-0.26) & 2.95 (-0.16) \\
5 & + PSF template & 98.60 (+0.17) & 0.70 (-0.05) & 2.57 (-0.37) \\
\hline
8 & Raw amp & 98.76 & 1.09 & 1.35 \\
8 & + EMCCD LL & 99.36 (+0.60) & 0.54 (-0.55) & 0.72 (-0.63) \\
8 & + Denoised amp & 99.54 (+0.18) & 0.32 (-0.21) & 0.57 (-0.15) \\
8 & + PSF template & 99.60 (+0.06) & 0.31 (-0.01) & 0.47 (-0.10) \\
\hline
10 & Raw amp & 99.44 & 0.52 & 0.59 \\
10 & + EMCCD LL & 99.72 (+0.28) & 0.25 (-0.27) & 0.30 (-0.29) \\
10 & + Denoised amp & 99.81 (+0.09) & 0.12 (-0.12) & 0.24 (-0.06) \\
10 & + PSF template & 99.83 (+0.02) & 0.12 (0.00) & 0.21 (-0.04) \\
\hline
12 & Raw amp & 99.74 & 0.43 & 0.20 \\
12 & + EMCCD LL & 99.90 (+0.16) & 0.13 (-0.30) & 0.10 (-0.10) \\
12 & + Denoised amp & 99.90 (0.00) & 0.15 (+0.03) & 0.08 (-0.02) \\
12 & + PSF template & 99.92 (+0.02) & 0.14 (-0.01) & 0.06 (-0.02) \\
\hline
14 & Raw amp & 99.88 & 0.21 & 0.09 \\
14 & + EMCCD LL & 99.94 (+0.06) & 0.07 (-0.14) & 0.06 (-0.03) \\
14 & + Denoised amp & 99.97 (+0.03) & 0.01 (-0.06) & 0.04 (-0.02) \\
14 & + PSF template & 99.97 (0.00) & 0.01 (0.00) & 0.03 (0.00) \\
\hline
16 & Raw amp & 99.91 & 0.04 & 0.11 \\
16 & + EMCCD LL & 99.94 (+0.03) & 0.05 (+0.01) & 0.06 (-0.05) \\
16 & + Denoised amp & 99.96 (+0.02) & 0.02 (-0.03) & 0.05 (-0.01) \\
16 & + PSF template & 99.97 (+0.01) & 0.02 (0.00) & 0.04 (-0.01) \\
\hline
20 & Raw amp & 99.95 & 0.05 & 0.05 \\
20 & + EMCCD LL & 99.98 (+0.03) & 0.04 (-0.01) & 0.01 (-0.04) \\
20 & + Denoised amp & 99.98 (0.00) & 0.03 (-0.01) & 0.01 (0.00) \\
20 & + PSF template & 99.98 (0.00) & 0.04 (+0.01) & 0.01 (-0.01) \\
\end{tabular}
\end{ruledtabular}
\end{table*}

\subsection{Additional candidate features}
\label{app:ablation_candidates}

We also tested the three additional candidate descriptor groups mentioned in
Sec.~III. Here we define their implemented forms and evaluate each group by
adding it separately to the selected four-feature classifier. This tests its
marginal contribution, rather than its predictive power in isolation. Let \(R(p)=I^{\mathrm{raw}}(p)-B^{\mathrm{raw}}(p)\) denote a
background-subtracted ROI in ADU, suppressing the site and shot indices.
Let \(\mathcal C\) be its central \(3\times3\) pixels and
\(\mathcal O=\mathcal R\setminus\mathcal C\) the outer 16 pixels of the
\(5\times5\) ROI.

\paragraph{\textbf{Center concentration.}}
The fraction of positive background-subtracted signal in the central pixels is
\begin{equation}
c_{\mathrm{center}}=
\frac{\sum_{p\in\mathcal C} R_+(p)}
     {\sum_{p\in\mathcal R} R_+(p)},
\qquad R_+(p)=\max[R(p),0].
\end{equation} Clipping the negative values avoids
cancellation in low-signal ROIs. This scalar describes spatial concentration.

\paragraph{\textbf{Learned noise score.}}
The second candidate uses the latent representation of the trained
Noise2NoiseFlow noise model. Raw images and the mean background are normalized
as \(\widetilde I\)
and \(\widetilde B\)
with the offsets and scale in ADU. The normalized images are replicated into
the two input channels \(c\) used by the saved noise model. Passing
\(\widetilde I-\widetilde B\) through its forward flow gives latent values
\(z_c(p)\). The ROI score is
\begin{equation}
a_{\mathrm{NF}}=-\frac{1}{2|\mathcal R|}
\sum_{p\in\mathcal R}\sum_{c=1}^{2}z_c(p)^2.
\end{equation}
A higher score indicates greater compatibility with the learned background
noise.

\paragraph{\textbf{PSF residual structure.}}
Using the same nonnegative, unit-sum empirical template \(T(p)\) as the PSF
correlation feature, we approximate the background-subtracted ROI by
\(R(p)\simeq A T(p)\), keeping the template shape fixed and fitting only
its amplitude. We choose the unconstrained amplitude by unweighted least squares,
\[
\widehat A=\operatorname*{arg\,min}_{A\in\mathbb R}
\sum_{p\in\mathcal R}[R(p)-A T(p)]^2.
\]
Setting the derivative with respect to \(A\) to zero gives
\(\sum_{p\in\mathcal R}T(p)[R(p)-\widehat A T(p)]=0\), yielding the
fitted amplitude and residual,
\begin{equation}
\widehat A=\frac{\sum_{p\in\mathcal R}R(p)T(p)}
{\sum_{p\in\mathcal R}T(p)^2},
\qquad r(p)=R(p)-\widehat A T(p).
\end{equation}
The residual group consists of three scalars,
\begin{align}
E_{\mathrm{center}}&=\sum_{p\in\mathcal C}r(p)^2,\\
E_{\mathrm{outer}}&=\sum_{p\in\mathcal O}r(p)^2,\\
q_{\mathrm{res}}&=\frac{\sum_{p\in\mathcal C}r(p)-\sum_{p\in\mathcal O}r(p)}
{\sqrt{\sum_{p\in\mathcal R}R(p)^2}},
\end{align}
These quantities describe residual
energy in the center and outer pixels and the signed center-to-outer imbalance.

Table~\ref{tab:supp_extra_features} reports the matched changes relative to
4D-LR. The largest observed fidelity increase is \(0.025\) percentage points,
from adding PSF residual structure at \(4~\mathrm{ms}\). All reported 95\%
intervals for the fidelity change include zero, and no candidate gives a
consistent improvement across measurement times. These results support
retaining the four-feature representation for the present linear classifier,
without ruling out benefits with other classifiers or datasets.

\begin{table*}[!ht]
\caption{
\label{tab:supp_extra_features}
\textbf{Additional candidate features added to the selected 4D-LR classifier.}
Each row adds only the indicated scalar or descriptor group to the same four
features. Changes are defined as augmented minus baseline performance:
\(\Delta F=F_{\mathrm{aug}}-F_{\mathrm{4D}}\),
\(\Delta\mathrm{FPR}=\mathrm{FPR}_{\mathrm{aug}}-\mathrm{FPR}_{\mathrm{4D}}\),
and \(\Delta\mathrm{FNR}=\mathrm{FNR}_{\mathrm{aug}}-\mathrm{FNR}_{\mathrm{4D}}\).
All values are in percentage points; positive \(\Delta F\) indicates improvement,
whereas positive rate changes indicate increased errors.
Brackets give paired 95\% percentile confidence intervals for \(\Delta F\),
using 5,000 frame-bootstrap draws with the same sampled frames for both
classifiers. The baseline out-of-fold decisions exactly match those used in
the main analysis at all eight measurement times.
}
\begin{ruledtabular}
\small
\begin{tabular}{clccc}
Exp. (ms) & Added candidate & \(\Delta F\) [95\% CI] & \(\Delta\)FPR & \(\Delta\)FNR \\
\hline
4 & Center concentration & 0.000 [-0.050, +0.050] & +0.117 & -0.029 \\
4 & Learned noise score & +0.006 [-0.013, +0.025] & -0.021 & -0.003 \\
4 & PSF residual structure & +0.025 [-0.011, +0.061] & -0.032 & -0.024 \\
\hline
5 & Center concentration & +0.014 [-0.032, +0.059] & +0.022 & -0.073 \\
5 & Learned noise score & 0.000 [-0.045, +0.050] & +0.022 & -0.036 \\
5 & PSF residual structure & -0.005 [-0.036, +0.023] & -0.007 & +0.024 \\
\hline
8 & Center concentration & -0.004 [-0.029, +0.018] & -0.009 & +0.013 \\
8 & Learned noise score & +0.004 [-0.007, +0.018] & +0.009 & -0.013 \\
8 & PSF residual structure & +0.015 [-0.004, +0.037] & -0.017 & -0.013 \\
\hline
10 & Center concentration & -0.004 [-0.018, +0.007] & 0.000 & +0.006 \\
10 & Learned noise score & +0.004 [0.000, +0.011] & 0.000 & -0.006 \\
10 & PSF residual structure & +0.004 [-0.007, +0.014] & -0.009 & 0.000 \\
\hline
12 & Center concentration & 0.000 [0.000, 0.000] & 0.000 & 0.000 \\
12 & Learned noise score & -0.007 [-0.018, 0.000] & +0.013 & +0.005 \\
12 & PSF residual structure & 0.000 [-0.022, +0.018] & +0.013 & -0.005 \\
\hline
14 & Center concentration & 0.000 [0.000, 0.000] & 0.000 & 0.000 \\
14 & Learned noise score & 0.000 [0.000, 0.000] & 0.000 & 0.000 \\
14 & PSF residual structure & 0.000 [0.000, 0.000] & 0.000 & 0.000 \\
\hline
16 & Center concentration & +0.004 [0.000, +0.011] & 0.000 & -0.005 \\
16 & Learned noise score & +0.004 [0.000, +0.011] & 0.000 & -0.005 \\
16 & PSF residual structure & -0.004 [-0.011, 0.000] & 0.000 & +0.005 \\
\hline
20 & Center concentration & -0.004 [-0.011, 0.000] & 0.000 & +0.006 \\
20 & Learned noise score & 0.000 [0.000, 0.000] & 0.000 & 0.000 \\
20 & PSF residual structure & -0.004 [-0.011, 0.000] & +0.009 & 0.000 \\
\end{tabular}
\end{ruledtabular}
\end{table*}

\clearpage
\section{Denoising Architecture}
\label{app:denoising}
Our denoising model is based on the Noise2NoiseFlow framework of
Maleky \textit{et al.}~\cite{maleky22}, which jointly trains an image denoiser and a
noise model using pairs of noisy images, without requiring any clean images for training.
The original framework was designed to accommodate variations in noise characteristics across different camera sensors and ISO settings. In contrast, our dataset consists exclusively of single-channel EMCCD fluorescence images acquired using the same Andor iXon Ultra 897 camera at a fixed EM gain of 300. We therefore simplified the original
architecture by removing the camera/ISO conditioning and the signal-dependent
camera-noise layers, while keeping the central idea of learning from paired
noisy observations.

For each exposure time, two consecutively acquired readout images were
used as a noisy image pair. These images contain the same nominal atom
configuration but independent camera-noise distributions. The denoiser
maps one noisy image to a denoised estimate, while the residual between the
denoised estimate and the paired noisy observation is evaluated by a 
normalizing-flow noise model.

The denoiser itself is a single-channel DnCNN~\cite{zhang2017dncnn} with
25 convolutional layers and one-channel input and output. While the original Noise2NoiseFlow implementation used a 9-layer DnCNN, we
tested the denoiser depth and
a 25-layer DnCNN gave the best validation performance on our EMCCD fluorescence
data.

The residual noise model uses a simplified unconditional flow. The flow
architecture is
\[
\begin{aligned}
&\mathrm{squeeze}
\,|\,
\mathrm{unc}
\,|\,
\mathrm{unc}
\,|\,
\mathrm{gain}
\,|\,
\mathrm{unc}
\,|\,
\mathrm{unc}
\\
&\quad|\,
\mathrm{gain}
\,|\,
\mathrm{unc}
\,|\,
\mathrm{unc}
\,|\,
\mathrm{unsqueeze}.
\end{aligned}
\]
The \textit{squeeze} operation rearranges a one-channel image into four channels,
which makes channel splitting possible in the following layers. Each
\textit{unc} block consists of an invertible \(1\times1\) convolution
followed by an unconditional affine-coupling layer. The role of these blocks is
to transform a simple gaussian distribution into a more complicated residual-noise
distribution through a sequence of invertible changes of variables. Stacking several such blocks allows the flow to represent
non-Gaussian and asymmetric EMCCD residual statistics that are not captured by a
single Gaussian model.
The \textit{gain} layers adjust
the overall scale of the modeled residual distribution.

The training objective follows the Noise2NoiseFlow form,
\[
\mathcal{L}
=
\mathcal{L}_{\mathrm{nm}}(\tilde I_1,\tilde I_2)
+
\lambda
\mathcal{L}_{\mathrm{dn}}(\tilde I_1,\tilde I_2),
\]
where \(\tilde I_1\) and \(\tilde I_2\) are two consecutive noisy atom images. The noise-model term
\(\mathcal{L}_{\mathrm{nm}}\) is the negative log-likelihood of the image
residual, defined as the difference between a noisy image and its
corresponding denoised image.
The denoising term \(\mathcal{L}_{\mathrm{dn}}\) compares the denoised
prediction obtained from one noisy image with the other independently acquired
noisy image of the same underlying scene. This cross-sample formulation enables
the denoiser to be trained without requiring a clean target image. The
coefficient \(\lambda\) controls the relative weight of the denoising term. We used the same loss weight as Ref.~\cite{maleky22},
$\lambda = 2^{18}$.
The model was optimized with Adam using
\(\beta_1=0.9\), \(\beta_2=0.999\), \(\varepsilon=10^{-8}\), and a constant
learning rate of \(10^{-3}\). The batch size was 8.

Each exposure time was trained independently. For each exposure time, the raw EMCCD
intensity was linearly normalized to \([0,1]\) using a separately chosen
\([v_{\min},v_{\max}]\) range. Training used \(64\times64\) grayscale images
with one patch per image. Training ran for at least 200 epochs and at most
500 epochs. Full validation was performed every 10 epochs, and after epoch 200
early stopping was triggered if the validation negative log-likelihood did not
improve by more than \(5\times10^{-4}\) for five consecutive
validation checks. 

\begin{figure}[!ht]
\centering
\includegraphics[
  width=\columnwidth,
  trim={3cm 10.5cm 2cm 3cm},
  clip
]{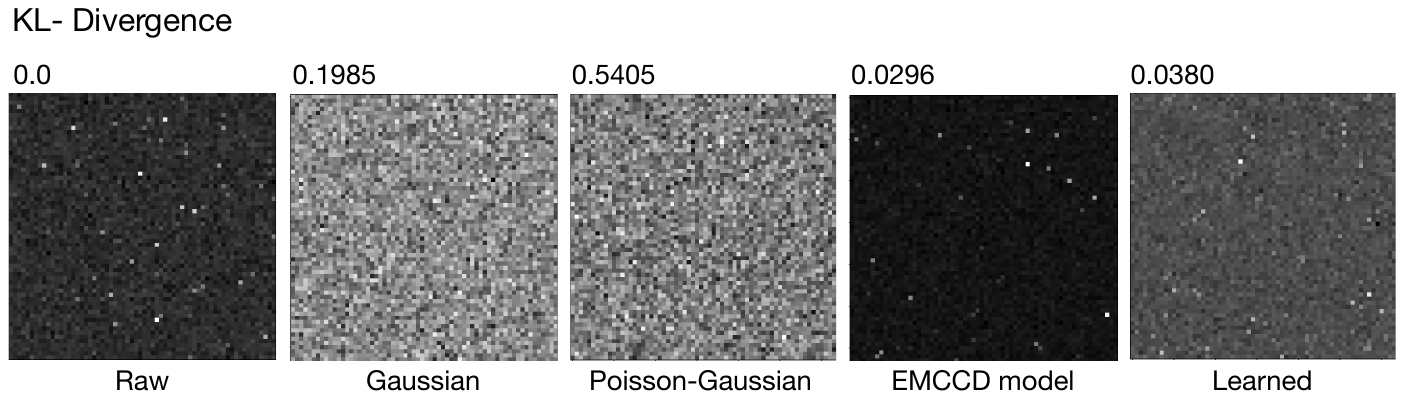}
\caption{\label{fig:noise_model}
\textbf{Comparison of learned and analytic EMCCD background-noise models.}
A real EMCCD background frame is compared with representative synthetic
background frames generated from four noise models: a Gaussian model, a
Poisson--Gaussian model, an analytic EMCCD model~\cite{martinezdorantes17}, and the learned residual
Noise2NoiseFlow~\cite{maleky22} model. The comparison was performed for the
\(t_{\mathrm{exp}}=8~\mathrm{ms}\) background data. For each noise model, we
generated synthetic background images and compared their pixel-value histogram
with the histogram of the measured background images. The same histogram bins
were used for all models so that the Kullback--Leibler divergence measures only
the difference in the generated pixel-value statistics. The number above each
image is \(D_{\mathrm{KL}}(P_{\mathrm{bg}}\Vert P_{\mathrm{model}})\), where
\(P_{\mathrm{bg}}\) is the measured background histogram and
\(P_{\mathrm{model}}\) is the histogram of the generated background images. The
real background frame is included as a visual reference and has zero divergence
by definition. Lower KL divergence indicates closer agreement with the measured
EMCCD background statistics.
}
\end{figure}

We evaluated whether the learned noise
model reproduces the measured EMCCD background statistics. For each noise model,
we generated synthetic background images and compared their pixel-value
histogram with the histogram of measured background images [Fig.\ref{fig:noise_model}]. The discrepancy was
quantified using the Kullback--Leibler divergence
\begin{equation}
D_{\mathrm{KL}}(P_{\mathrm{bg}}\Vert P_{\mathrm{model}})
=
\sum_b
P_{\mathrm{bg}}(b)
\log
\frac{P_{\mathrm{bg}}(b)}
     {P_{\mathrm{model}}(b)},
\end{equation}
where \(P_{\mathrm{bg}}\) is the empirical background pixel-value distribution,
\(P_{\mathrm{model}}\) is the distribution generated by a candidate noise model,
\(b\) indexes histogram bins. Lower
\(D_{\mathrm{KL}}\) indicates better agreement with the measured EMCCD
background statistics. Figure \ref{fig:noise_model} compares Gaussian,
Poisson--Gaussian, analytic EMCCD~\cite{martinezdorantes17}, and learned Noise2NoiseFlow background
models. The analytic EMCCD model gives the closest agreement with the measured
background distribution among the tested models
\((D_{\mathrm{KL}}=0.0296)\), while the learned Noise2NoiseFlow residual model
achieves a comparable value \((D_{\mathrm{KL}}=0.0380)\). This indicates that
the learned model captures much of the measured EMCCD background statistics.

If denoising introduces a spot resembling atomic fluorescence at an empty site,
the site may be misclassified as occupied. Such a spurious feature is an example
of image hallucination. Generative adversarial network (GAN)-based image restoration methods can improve perceptual realism
by matching the output image distribution, but this objective can trade off
against distortion fidelity and can generate structures not supported by the
input data \cite{blau18_perception_distortion,cohen18_hallucinate}. Similar concerns apply to GAN-like or image-translation-based denoising
approaches when the objective emphasizes visual plausibility rather than
statistical consistency with the measured data.

\begin{figure*}[!ht]
\centering
\includegraphics[width=\textwidth]{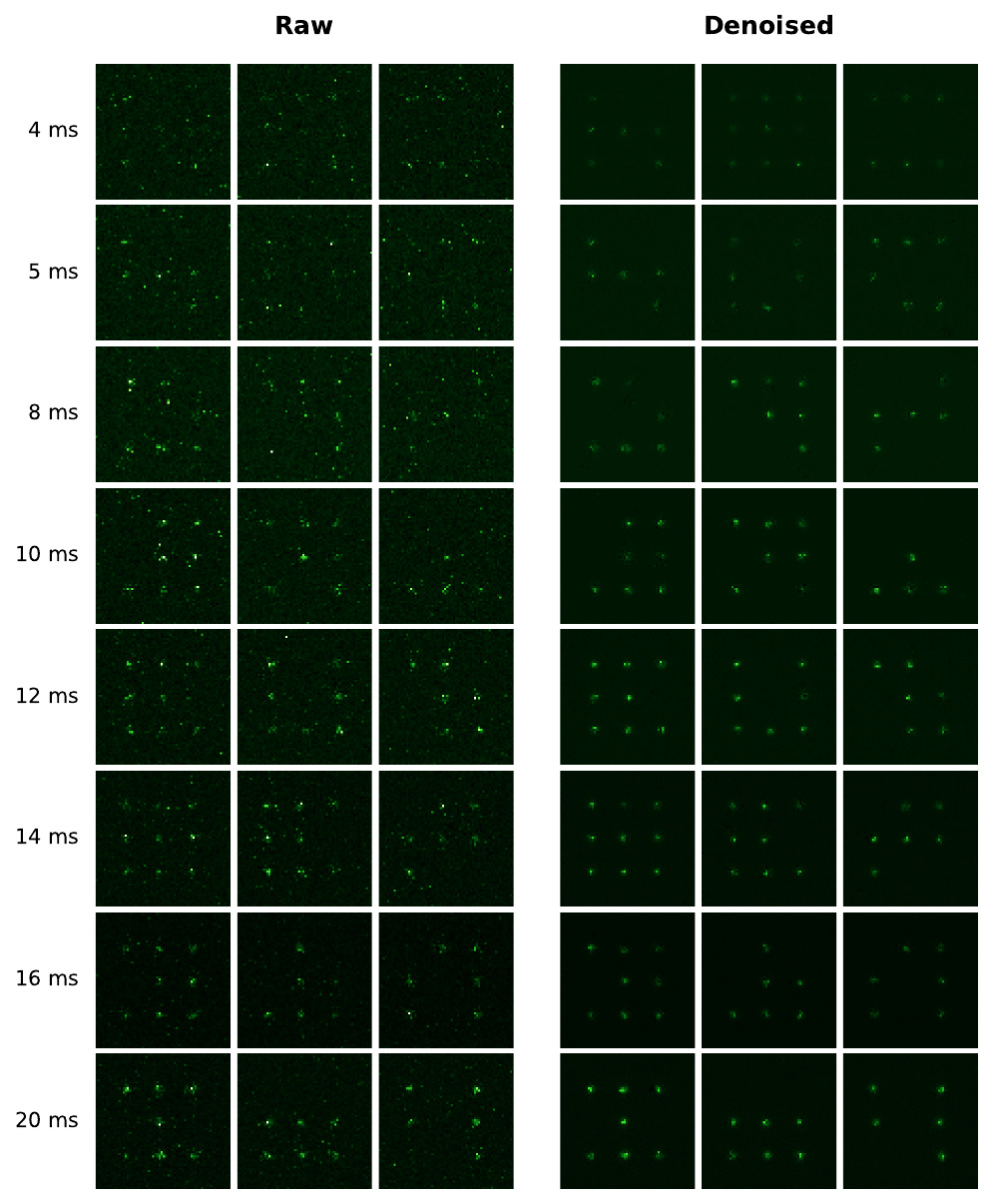}
\caption{\label{fig:wide}
\textbf{Gallery of raw and denoised fluorescence images.}
Samples of raw and denoised images are shown for all measurement
times \(t_{\mathrm{img}}=4,5,8,10,12,14,16,\) and \(20~\mathrm{ms}\). Within each row, the first three images are raw EMCCD images, and the last three images are the corresponding denoised outputs.
}
\end{figure*}

The employed denoising architecture is more conservative in this respect. The DnCNN
module enables to predict a residual image, and the normalizing-flow module models the statistics of
that residual.
Thus, the residual after denoising is encouraged to remain compatible with the
measured camera-noise distribution. If the denoiser were to create or remove an
atom-like feature, the corresponding residual would contain
signal-like components rather than background-like noise. The flow-based
residual likelihood and the background-statistics validation therefore provide
a safety check against such artifacts. Compared with adversarial image-generation objectives, the
DnCNN+Flow formulation is better aligned with conservative scientific
denoising, because it models the noise residual explicitly and can be tested
against measured background statistics. Figure~\ref{fig:wide} shows examples of raw fluorescence images and their
denoised counterparts at each exposure time.

\clearpage
\section{Rearrangement simulation}
\label{app:rearrange}

This appendix describes the Monte Carlo rearrangement simulation used to evaluate the effects of readout errors on defect-free atom array preparation. The simulation focuses on the asymmetry between false positives and false negatives during source assignment. A false positive can select an empty reservoir site as an atom source and directly create a target defect, whereas a false negative mainly reduces the available atom pool. We therefore use a simplified model that isolates readout-induced source-assignment errors, rather than modeling the detailed transport trajectory, collision constraints, or motion-control errors of a specific rearrangement algorithm~\cite{kim19_rearrangement, lin25_aiassembly}.

Each trial starts from an \(M\times M\) reservoir array with independent loading
probability \(p_{\mathrm{load}}\). For a given readout method and exposure time,
imaging survival is first sampled using \(P_{\mathrm{surv}}(t_{\mathrm{img}})\).
The measured false-negative rate \(p_{\mathrm{FN}}\) is applied to surviving atoms,
and the false-positive rate \(p_{\mathrm{FP}}\) to sites empty before imaging.
The rearrangement
algorithm then selects source atoms from the measured occupied sites and attempts
to fill a square target array.

For single-round rearrangement, the target is fixed at \(20\times20\) and
only one readout-and-move cycle is allowed. For multi-round rearrangement,
the target side in round \(k\) is \(L_k=\lfloor\sqrt{B}\rfloor\), where
\(B\) is the number of detected candidates retained after erasure filtering.
The trial is counted as successful only if the assembled array is defect-free
and has side length at least \(L_{\mathrm{floor}}=20\), within at most 20 rounds. After a failed round, the next round is initialized only
with atoms that are actually present in the attempted target region. Any unused atoms remaining in the reservoir are discarded and not reused. 

For erasure-aware source selection, low-confidence occupied-site predictions are
removed from the candidate source pool before rearrangement. This suppresses false source assignments caused by
false positives, but also discards some truly occupied atoms. The sweep over erasure budgets
therefore measures the tradeoff between removing unreliable source
assignments and preserving a sufficiently large atom pool.
The effective simulator inputs are
\begin{align}
 p_{\mathrm{FP}}^{\mathrm{eff}}
 &= \frac{\mathrm{FP}_{\mathrm{kept}}}{N_{\mathrm{empty,total}}}
 = p_{\mathrm{FP}}(1-p_{\mathrm{era}|F}),\\
 p_{\mathrm{FN}}^{\mathrm{eff}}
 &= \frac{\mathrm{FN}_{\mathrm{kept}}+N_{\mathrm{erased,atom}}}
 {N_{\mathrm{atom,total}}}
 = 1-(1-p_{\mathrm{FN}})(1-p_{\mathrm{era}|T}).
\end{align}
Here, the denominators include all reference-labeled atom-absent or atom-present ROI
outcomes before erasure filtering.
Algorithm~\ref{alg:rearrangement} shows this equivalent two-step thinning
using unfiltered error rates and conditional candidate-erasure probabilities. Sources are selected
uniformly without replacement from the retained candidates.
Each condition uses 10,000 trials and defect tolerance zero.
For erasure-aware methods, the lowest budget within one Monte Carlo standard
error of the maximum success rate is selected from the same sweep, with a
maximum budget of 25\%. The success of each simulated trial is evaluated from the known simulated
ground-truth occupation after rearrangement. This is a simulation oracle
criterion, rather than an experimentally implemented stopping measurement.

In the simulation, true-positive source candidates are sampled from atoms surviving imaging, and false-positive candidates from sites that were empty before imaging. We assume that loss-related false-positive assignments are already included in the empirical FPR. Therefore, atoms lost during imaging are not added as separate false-positive source candidates. No additional readout or refill is performed within a round.

\begin{figure*}[!ht]
\begin{minipage}{\textwidth}
\begin{algorithm}[H]
\caption{Monte Carlo rearrangement simulation}
\label{alg:rearrangement}
\begin{algorithmic}[1]
\Require Reservoir size \(M\), loading probability \(p_{\mathrm{load}}\),
readout error rates \(p_{\mathrm{FP}}\) and \(p_{\mathrm{FN}}\), conditional erasure probabilities
\(p_{\mathrm{era}|T}\) and \(p_{\mathrm{era}|F}\), imaging survival probability
\(P_{\mathrm{surv}}\), protocol mode, maximum rounds \(K\), fixed target side
\(L\), minimum target side \(L_{\mathrm{floor}}\), and defect tolerance \(\tau\).
\Statex \hspace{\algorithmicindent}
Here, \(p_{\mathrm{era}|T}\) is the probability that a true-positive candidate is
erased, and \(p_{\mathrm{era}|F}\) is the probability that a false-positive
candidate is erased. Without erasure-aware source selection, both probabilities are set
to zero.
\State \(A \sim \mathrm{Binomial}(M^2,p_{\mathrm{load}})\), \(S \leftarrow M^2\)
\Comment{\(A\): true atoms, \(S\): available sites}
\For{\(k=1,\ldots,K\)}
    \State \(A_{\mathrm{pre}} \leftarrow A\)
    \State \(A \leftarrow \mathrm{Binomial}(A_{\mathrm{pre}},P_{\mathrm{surv}})\)
    \Comment{surviving atoms after imaging}
    \State \(B_{\mathrm{TP}} \leftarrow \mathrm{Binomial}(A,1-p_{\mathrm{FN}})\)
    \Comment{surviving atoms detected as occupied}
    \State \(B_{\mathrm{FP}} \leftarrow \mathrm{Binomial}(S-A_{\mathrm{pre}},p_{\mathrm{FP}})\)
    \Comment{initially empty sites falsely detected as occupied}
    \State \(B_{\mathrm{TP}} \leftarrow \mathrm{Binomial}(B_{\mathrm{TP}},1-p_{\mathrm{era}|T})\)
    \Comment{true-positive candidates kept after erasure filtering}
    \State \(B_{\mathrm{FP}} \leftarrow \mathrm{Binomial}(B_{\mathrm{FP}},1-p_{\mathrm{era}|F})\)
    \Comment{false-positive candidates kept after erasure filtering}
    \State \(B \leftarrow B_{\mathrm{TP}}+B_{\mathrm{FP}}\)
    \If{protocol mode is single-round}
        \State \(L_k \leftarrow L\)
        \If{\(B < L^2\)}
            \State \Return failure
        \EndIf
    \Else
        \State \(L_k \leftarrow \lfloor\sqrt{B}\rfloor\)
        \If{\(L_k < L_{\mathrm{floor}}\)}
            \State \Return failure
        \EndIf
    \EndIf
    \State \(T \leftarrow \mathrm{Hypergeometric}(B,B_{\mathrm{TP}},L_k^2)\)
    \Comment{population \(B\), real candidates \(B_{\mathrm{TP}}\), draws \(L_k^2\)}
    \If{\((L_k^2-T)/L_k^2 \leq \tau\)}
        \State \Return success
    \Else
        \State \(A \leftarrow T\), \(S \leftarrow L_k^2\)
        \Comment{conservative cleanup: reuse only the attempted target region}
    \EndIf
\EndFor
\State \Return failure
\end{algorithmic}
\end{algorithm}
\end{minipage}
\end{figure*}

\clearpage
\twocolumngrid
\makeatletter
\def\@bibstyle{apsrev4-2-tenauthors}
\makeatother
\bibliography{apssamp}

\end{document}